\documentclass[journal, twocolumn,final]{IEEEtran}
\usepackage{amsfonts}
\usepackage{amssymb}
\usepackage{amsmath}

\usepackage{dsfont}
\usepackage{graphicx}
\usepackage{bm}
\usepackage{cite}
\usepackage{bigstrut}
\usepackage{float}
\usepackage{balance}
\usepackage{lipsum}
\usepackage{psfrag}
\usepackage{xcolor}
\usepackage{xfrac}
\usepackage{hyperref}

\usepackage[ruled,vlined]{algorithm2e}
\usepackage{tikz}
\usepackage{pgfplots}
\pgfplotsset{compat=1.15}
\usepackage{mathrsfs}
\usetikzlibrary{arrows}
\usetikzlibrary{decorations.pathreplacing}
\usetikzlibrary{fadings}
\usetikzlibrary{fit}

\newtheorem{lemma}{Lemma}

\newtheorem{remark}{Remark}
\usepackage{mathtools}
\usepackage{comment}
\SetKwInput{KwInput}{Input}                % Set the Input
\SetKwInput{KwOutput}{Output}

\begin{document}
\IEEEoverridecommandlockouts
\newcommand\norm[1]{\left\lVert#1\right\rVert}
\pgfplotsset{compat=1.17}
\usepgfplotslibrary{patchplots}   % ← enables bilinear patches

\title{Physical Layer Security with Artificial Noise in MIMO Pinching-Antenna Systems}
\author{Pigi P. Papanikolaou, Dimitrios Bozanis, Sotiris A. Tegos,~\IEEEmembership{Senior Member,~IEEE}, \\
Panagiotis D. Diamantoulakis,~\IEEEmembership{Senior Member,~IEEE}, Panagiotis Sarigiannidis,~\IEEEmembership{Member,~IEEE},  \\ and George K. Karagiannidis,~\IEEEmembership{Fellow,~IEEE}

%\IEEEauthorblockA{\IEEEauthorrefmark{1}Department of Electrical and Computer Engineering, Aristotle University of Thessaloniki, Greece\\}
%\IEEEauthorblockA{e-mail: \{pigipapa, dimimpoz, tegosoti, padiaman, geokarag\}@auth.gr}
\thanks{P. P. Papanikolaou, D. Bozanis, S. A. Tegos, P. D. Diamantoulakis and G. K. Karagiannidis are with the Department of Electrical and Computer Engineering, Aristotle University of Thessaloniki, 54124 Thessaloniki, Greece (e-mails: pigipapa@auth.gr, dimimpoz@auth.gr, tegosoti@auth.gr, padiaman@auth.gr, geokarag@auth.gr).}
\thanks{P. Sarigiannidis is with the Department of Electrical and Computer Engineering, University of Western Macedonia, 50100 Kozani, Greece (e-mail: psarigiannidis@uowm.gr).}
\thanks{Part of this work has been accepted for publication in 2025 IEEE Personal, Indoor and Mobile Radio Communications (PIMRC) \cite{our}.}
\vspace{-5mm}

%\thanks{The authors are with the Department of Electrical and Computer Engineering, Aristotle University of Thessaloniki, 54124 Thessaloniki, Greece (e-mails: pigipapa@ece.auth.gr, dimimpoz@ece.auth.gr, tegosoti@auth.gr, padiaman@auth.gr, geokarag@auth.gr).}
%\thanks{The work of G. K. Karagiannidis was implemented in the framework of HFRI call ``Basic research financing (horizontal support of all sciences)'' under the National Recovery and Resilience Plan ``Greece 2.0'' funded by the European Union - NextGenerationEU (HFRI Project Number: 15642).}
%\vspace{-.4cm}
}

\maketitle
%with the Wireless Communications and Information Processing Group (WCIP), Department of Electrical and Computer Engineering, Aristotle University of Thessaloniki and 

\begin{abstract}
As next-generation wireless networks emerge, security is becoming a critical performance metric. However, conventional multiple-input-multiple-output (MIMO) systems often suffer from severe path loss and are vulnerable to nearby eavesdroppers due to their fixed-antenna configurations. Pinching-antenna systems (PASs) offer a promising alternative, leveraging reconfigurable pinching antennas (PAs) positioned along low-loss dielectric waveguides to enhance channel conditions and dynamically mitigate security threats. In this paper, we propose an artificial noise (AN)-aided beamforming framework for the PAS downlink that maximizes the secrecy rate (SR) by jointly optimizing the information beams, the AN covariance, and the PA positions. We examine both perfect and imperfect channel state information (CSI) for the eavesdropper's channel. For the latter, location errors are mapped via a Jacobian into an ellipsoidal channel uncertainty set to accurately formulate the problem. We derive a closed-form solution for the single-waveguide scenario, yielding the optimal PA location and an information/AN power-splitting rule. For multiple waveguides and users, we develop a deep neural network (DNN)-aided joint optimizer that outputs beams, AN, and PA placements. Numerical results demonstrate that the proposed scheme improves SR consistently over PAS baselines in single- and multi-user settings under both perfect and imperfect CSI.
\end{abstract}


\section{Introduction}\label{sec:Intro}

\subsection{Background and Motivation}

In recent decades, wireless communication systems have advanced significantly, driven by the relentless pursuit of higher data rates, improved reliability, and enhanced security. Among various technologies, multiple-input-multiple-output (MIMO) systems have played a crucial role by introducing spatial degrees of freedom (DoFs), enabling beamforming techniques, and significantly increasing spectral efficiency \cite{MIMO2}. However, conventional MIMO systems typically rely on fixed antenna configurations, limiting their ability to respond dynamically to changes in the propagation environment, such as obstacles, user mobility, or evolving network requirements \cite{tyro}. In this rapidly growing environment of sensitive data and the increased number of potential eavesdroppers, security remains an urgent challenge for future sixth-generation (6G) networks \cite{6g}. While encryption and decryption have long served as fundamental pillars of information security, their high computational complexity and intricate key management increasingly struggle to keep pace with the ever-expanding data demands of next-generation networks \cite{springer}. To address this issue, physical layer security (PLS) has been proposed as a highly promising technology, garnering significant attention in recent years \cite{PLS, PLS_2}.

Recently, the concept of dynamic wireless channel reconfiguration has emerged, leveraging technologies such as reconfigurable intelligent surfaces (RISs) \cite{RIS}, movable antennas \cite{MOVE}, and fluid antennas \cite{fluid}. %Despite their advantages, these approaches still face limitations, including restricted reconfiguration capabilities, limited movement ranges, and challenges in dealing with severe large-scale path loss and line-of-sight (LoS) blockage.
These technologies improve the quality of legitimate channels while suppressing eavesdropper reception by dynamically reshaping propagation environments, thereby boosting the secrecy rate (SR) \cite{Zhang_2021, xia, ghadi, ding_the_second}. These reconfigurable architectures have attracted significant attention and demonstrated notable performance improvements. However, they are limited in their ability to combat large-scale path loss. Movable and fluid antennas typically operate over only a few wavelengths, which primarily mitigates small-scale fading. Furthermore, while RISs can synthesize virtual line-of-sight (LoS) links, they suffer from double attenuation, which leads to higher path loss \cite{RIS_2}. Furthermore, many flexible-antenna systems have limited reconfiguration capabilities and movement ranges, making it difficult to adjust the number of active elements to meet practical communication demands. These limitations may hinder their use in some high-frequency or longer-range settings, suggesting the need for more adaptive designs.

In response to these challenges, pinching-antenna systems (PASs) are emerging as a promising flexible-antenna solution. Originally demonstrated by NTT DOCOMO \cite{DOCOMO}, PASs leverage low-loss dielectric waveguides and deploy small dielectric particles, i.e., pinching antennas (PAs). This enables significant reconfiguration capabilities, including adjusting antenna positions, effectively mitigating path loss by establishing strong LoS links and improving spatial channel control with minimal complexity and cost \cite{10909665}. As the electromagnetic signals propagating along the waveguide leak from multiple antennas, PAs share certain characteristics with leaky-wave antennas \cite{yang}, which have recently been explored for applications such as holographic MIMO systems \cite{holographic}. Unlike leaky-wave systems, where element spacing is tied to the wavelength and thus constrained in its ability to mitigate large-scale path loss, PAs can be activated at subwavelength intervals along a dielectric waveguide.

Recent research on PASs has established both their physical advantages and their algorithmic potential. Foundational works position PAS as a flexible antenna architecture that can create strong LoS links and mitigate large-scale path loss, with multi-pinch/multi-waveguide variants linking to non-orthogonal multiple access (NOMA) and multiple-input single-output (MISO) formulations \cite{ding}. Based on this, \cite{arxigos} maximized the downlink rate by optimizing PA locations using a two-stage algorithm, significantly outperforming the fixed-antenna baselines. In \cite{wang}, the authors built practical models for PASs and propose methods that jointly control the transmit signals and the positions of the pinching antennas, both of which showed notable performance gains. For multi-user MIMO, \cite{bereyhi} developed a hybrid beamforming framework and a low complexity fractional-programming based algorithm that jointly tunes precoders and PA positions to boost a weighted sum-rate metric. In \cite{ody}, the authors modeled PA selection as a quadratic fractional program and train multi-layer perceptron (MLP) and graph neural network (GNN)-based models to choose the best subset of preinstalled PAs, achieving near-optimal rates. Moreover, PAS were proposed for integrated sensing and communication in \cite{bozanis}, where the Cramér–Rao bounds for both range and angle estimation were derived, offering significant gains in sensing capabilities, but leaving security and joint waveform design untouched.

Although this fundamental research has provided important insights into PASs, it has primarily focused on maximizing data rates without considering the inherent security risks. Due to the broadcast nature of wireless communications, confidential transmissions are particularly vulnerable to eavesdropping, motivating the development of PLS techniques. Recent studies have begun to explore the integration of PLS into PAS. In \cite{huawai}, the authors proposed fractional programming and gradient-based algorithms to design the optimal beamforming matrices that maximize SRs, though they still depend on perfect CSI. In contrast, in \cite{brasil}, expressions for key security metrics such as secrecy failure probability and secrecy capacity were derived, highlighting the performance gains achievable through the strategic use of PAs. The study in \cite{wang_the_second} provided closed-form secrecy metrics for a single-PA PAS under perfect CSI, showing how PA placement affects secrecy. Furthermore, a PAS for covert communication was studied in \cite{covert}, where PA placement was optimized to hide transmissions from an eavesdropper while maintaining throughput. However, none of these works considered multi-user and joint AN/beamforming design. 
% Finally, in our work \cite{our}, we introduced a joint AN, beamforming, PA-placement approach for PAS, with one PA per waveguide, for a single-user scenario with perfect CSI of the eavesdropper's channel.

\subsection{Contribution}

Existing works on PASs have largely focused on optimizing data rates under ideal CSI, or proposing single-user secrecy heuristics solutions that still require per-instance iterations. These solutions used complex and alternating algorithms that decrease performance and do not jointly exploit baseband beamforming, artificial noise (AN), and PA positioning. Robust treatments of uncertain Eve locations are also limited. In this paper, we develop a secrecy‐aware MIMO PAS for multi-user, multi-eavesdropper, multi-PA, multi-waveguide downlink. In this context, we jointly design the beamformers, the AN covariance, and the PA positions to maximize the worst-case SR. Two CSI cases are considered: (i) perfect CSI and (ii) imperfect eavesdropper's location knowledge, where location errors are mapped via a Jacobian linearization into an ellipsoidal channel-uncertainty set for an accurate error model. 
% To handle the arising highly non-convex problem, we design a deep neural network (DNN)-aided optimizer that jointly produces the beamforming matrices, the AN covariance matrix, and the PA position matrix, while enforcing power and spacing constraints through appropriate penalty terms. 
% We also derive a closed-form single-waveguide solution that offers significant geometric design insights, while extensive simulations quantify the gain under various scenarios and parameters.

More specifically, the contributions of our work are summarized as follows:
\begin{itemize}
    \item We formulate SR maximization problems for MIMO PAS consisting of multiple users, multiple eavesdroppers, multiple waveguides, and multiple PAs per waveguide, jointly optimizing the beamforming, AN covariance, and PA positions matrices. The framework covers both perfect and imperfect eavesdropper's CSI knowledge.
    \item For a single-waveguide deployment, we derive a closed-form solution that provides the optimal PA position and yields a simple power-splitting rule between information and AN.
    \item For the general multi-waveguide case, we design a deep neural network (DNN)-aided optimizer that jointly produces the beamforming matrices, the AN covariance matrix, and the PA position matrix to handle the arising highly non-convex problem.  Power and spacing constraints are enforced via thoroughly designed penalty terms, 
    %while the imperfect-CSI case is faced for the worst-case Eve-SINR bounds within an uncertainty ellipsoid.
    while under imperfect-CSI, the worst-case SR is maximized, where Eves' channel uncertainty is modeled by an ellipsoid and handled via S-procedure.
    \item We provide extensive numerical results, including convergence traces, SR–power curves, uncertainty-sweep cumulative distribution functions (CDFs), and spatial heatmaps, to validate the convergence and effectiveness of the proposed closed-form and DNN-aided schemes and the gains over state-of-the-art benchmarks.
\end{itemize}

\subsection{Structure}

This paper is organized as follows. Section \ref{sec:SysMod} presents the PAS downlink multiple-waveguides, multi-user, multi-eavesdropper PAS, the channel and signal models, and the SR metric. In Section \ref{sec3}, the perfect CSI scenario is studied, where an SR maximization problem is formulated. We derive a closed-form solution for the single-waveguide deployment, while we introduce a DNN-aided joint optimizer for the multi-waveguide/multi-user setting. In Section \ref{sec4}, the imperfect eavesdropper's channel knowledge is addressed by mapping location errors to an ellipsoidal channel-uncertainty set 
%and enforcing worst-case SINR bounds via an S-procedure LMI embedded in the learning objective.
and by enforcing worst-case (largest) Eve SINR over this set via an S-procedure LMI embedded in the learning objective.
Finally, Section \ref{sec:Num} provides numerical results for various setups and parameters, while Section \ref{sec:Conc} concludes the paper.

\section{System Model}\label{sec:SysMod}

\begin{figure}
    \centering
    % \includegraphics[height=5.3cm]
    % \vspace{-4mm}
    \includegraphics[width=1\columnwidth]{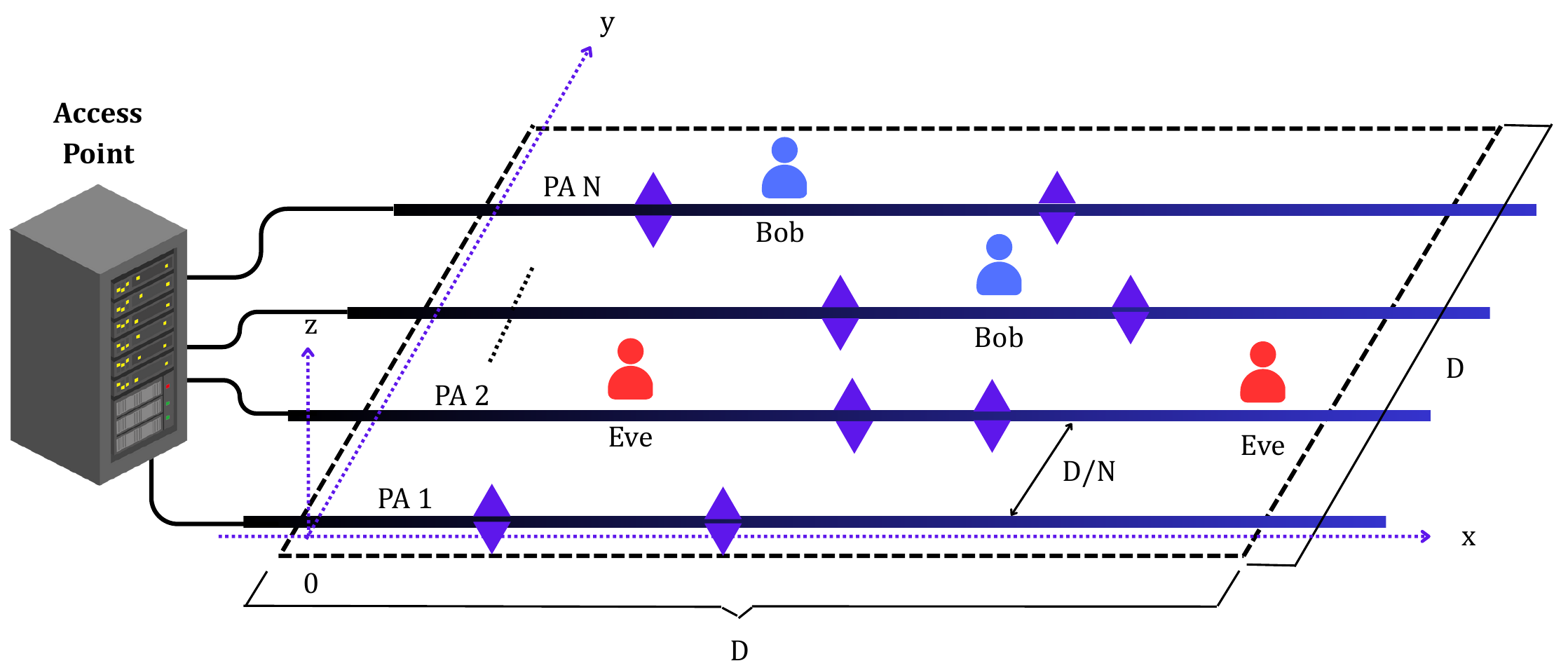}
    \vspace{-2mm}
    \caption{Illustration of a downlink MIMO PAS with multiple waveguides, PAs, eavesdroppers and users.}
    \vspace{-4mm}
    \label{fig:sys_model}
\end{figure}

In this work, we consider a secure downlink transmission framework based on PAs, as illustrated in Fig. \ref{fig:sys_model}. The base station (BS) is connected with $N$ waveguides, with one RF-chain corresponding to each of them. In each waveguide, by deploying small dielectric particles whose positions can be dynamically adjusted along the waveguide, multiple PAs are activated. Unlike the single PA per waveguide assumption that is common in literature \cite{wang,bereyhi,huawai}, this multi-antenna configuration provides additional spatial degrees of freedom. %Thus, advanced beamforming techniques are facilitated, and the potential for spatial reconfiguration and secure transmissions is further improved. 

Let $\mathcal{N}=\{1, 2, ..., N\}$ denote the set of waveguides, where the $n$-th waveguide is equipped with a set of PAs $\mathcal{M}_n=\{1, 2, ..., M_n\}$. However, in this work without loss of generality, we consider $M_n=M$, $\forall n \in \mathcal{N}$. At the BS, the confidential information is first processed via baseband beamforming, where it is superimposed with AN to enhance security, and then routed to the RF chains for up-conversion and transmission. 
%This reconfigurability enables the BS to fine-tune the spatial transmission pattern, thereby enhancing secrecy by strengthening the legitimate link and hindering eavesdropping attempts. 
Moreover, without loss of generality, all waveguides are assumed to be aligned parallel to the x-axis, uniformly spaced, and mounted at a fixed height $d$. In our scenario, we have $I$ legitimate users (Bob) and $K$ eavesdroppers (Eve), which are equipped with a single antenna and are distributed within a square region on the $x-y$ plane with a side length of $D$, with positions $\boldsymbol{\psi}_{B, i} = [x_{b,i}, y_{b, i}, 0]$, $i\in \mathcal{I}$ and $\boldsymbol{\psi}_{E, k} = [x_{e, k}, y_{e, k}, 0]$, $k\in\mathcal{K}$, respectively, where $\mathcal{I}=\{1, 2, ..., I\}$ and $\mathcal{K}=\{1, 2, ..., K\}$. Furthermore, the length of each waveguide is assumed to be $D$ to ensure complete coverage of the area. The $y$-axis coordinate of the $n$-th waveguide is given as $\tilde{y}_{n,0}=(n-1)D/N$, thus the position of the $m$-th PA on the $n$-th waveguide can be expressed as $\boldsymbol{\tilde{\psi}}_{n,m}=[\tilde{x}_{n,m},\:\tilde{y}_{n,m},\:d]$, where $\tilde{y}_{n,m}=\tilde{y}_{n,0}$.

The in-waveguide channel coefficient from the feed point of the $n$-th waveguide to the $m$-th PA is given by
\begin{equation}\label{channel1}
h^{(1)}_{n,m} = \frac{1}{\sqrt{M}}e^{ - j\frac{2\pi}{\lambda_g}\lVert \tilde{\boldsymbol{\psi}}_{n,0}-\tilde{\boldsymbol{\psi}}_{n,m} \rVert},
\end{equation}
where $\tilde{\boldsymbol{\psi}}_{n,0} = [\tilde{x}_{n,0}, \tilde{y}_{n,0}, d]$ denotes the location of the feed point of the $n$-th waveguide, and $\lambda_g = \frac{\lambda}{n_{\mathrm{eff}}}$ denotes the guided wavelength with $n_{\mathrm{eff}}$ being the effective refractive index of a dielectric waveguide and $\lambda$ is the wavelength
 in free space. 
Furthermore, the wireless channel coefficient between the $m$-th PA on the $n$-th waveguide and the $i$-th Bob can be modeled as 
\begin{equation}\label{channel2}
h^{(2)}_{b_i,n,m} = \frac{\eta^{\frac{1}{2}}e^{-j\frac{2\pi}{\lambda}\lVert \boldsymbol{\psi}_{B_i}-\tilde{\boldsymbol{\psi}}_{n,m} \rVert }}{\lVert \boldsymbol{\psi}_{B_i}-\tilde{\boldsymbol{\psi}}_{n,m} \rVert},
\end{equation}
where $\eta=\frac{c^2}{16\pi^2 f_c^2}$ is a constant with $c$ denoting the speed of light and $f_c$ is the carrier frequency. 

Hence, the legitimate channels to all Bobs are given by 
\begin{equation}\label{channel}
\boldsymbol{H}_{B} = \boldsymbol{H}_B^{(2)} \boldsymbol{H}^{(1)} \in \mathbb{C}^{I\times N} ,
\end{equation}
where $\boldsymbol{H}_B^{(2)} \in \mathbb{C}^{I\times L}$ and the elements of the matrix are $\left( \boldsymbol{H}_B^{(2)}\right)_{i,(n,m)}=h^{(2)}_{b_i,n,m}$, where $L= \sum_{n\in \mathcal{N}} M_n$ and index $(n,m)$ indicates function $(n-1)M+m$. Also, $\boldsymbol{H}^{(1)}\in  \mathbb{C}^{L\times N}$ and the elements of this matrix are given as $\left( \boldsymbol{H}^{(1)}\right)_{(n,m),n}=h^{(1)}_{n,m}$ and $\left( \boldsymbol{H}^{(1)}\right)_{(n',m),n}=0$, if $n' \neq n$, i.e., $\boldsymbol{H}^{(1)}$ is a block diagonal matrix with $M\times 1$ blocks. The communication channel matrix $\boldsymbol{H}_{B}$ is assumed to be known to the BS and its entries are independently distributed. As observed from \eqref{channel}, the channel in the PAS is affected by free-space path loss and two phase shift terms, where the first is caused by the signal propagation in the free space and the second is caused from the in-waveguide propagation. By multiplying $\boldsymbol{H}^{(1)}$ and $\boldsymbol{H}_B^{(2)}$, the equivalent per-waveguide channel between the $n$-th waveguide and $i$-th Bob becomes 
\begin{equation}
    \left( \boldsymbol{H}_B\right)_{i,n}=\displaystyle\sum_{\substack{m=1}}^{M} h^{(2)}_{b_i,n,m} h^{(1)}_{n,m},
\end{equation}
% where the factor $1/\sqrt{M}$ normalizes the per-waveguide channels so that the total transmit power across the $M$ RF chains remains fixed when $M$ varies. 
thus we define the per-waveguide channel vector for Bob $i$ as $\boldsymbol{h}_{B,i}=[\left( \boldsymbol{H}_B\right)_{i,1}, ..., \left( \boldsymbol{H}_B\right)_{i,N}]^T\in \mathbb{C}^{N\times 1}$.

The transmitted signal 
%$\boldsymbol{X}\in \mathbb{C}^{L\times 1}$ 
$\boldsymbol{x}\in \mathbb{C}^{N\times 1}$
can be written as
%\begin{equation}
%    \boldsymbol{X} = \boldsymbol{W} \boldsymbol{S} + \boldsymbol{M},
%\end{equation}
\begin{equation}
    \boldsymbol{x} = \boldsymbol{W} \boldsymbol{s} + \boldsymbol{m},
\end{equation}
with $\boldsymbol{W} = [\boldsymbol{w}_1, \boldsymbol{w}_2, ..., \boldsymbol{w}_I]\in \mathbb{C}^{N \times I}$ being the beamforming matrix, where $\boldsymbol{w}_i \in \mathbb{C}^{N \times 1}$ is the beamforming vector for Bob $i\in \mathcal{I}$, while each element of $\boldsymbol{s}\in \mathbb{C}^{I\times 1}$ denotes the $i$-th unit-power data stream intended to the $i$-th Bob, and $\boldsymbol{m}\in \mathbb{C}^{N \times 1}$ is the AN vector generated by the transmitter to create interference to potential eavesdroppers. We assume $\boldsymbol{m} \sim \mathcal{CN}(\boldsymbol{0}, \boldsymbol{R}_m)$, where $\boldsymbol{R}_m \in \mathbb{C}^{N\times N} \succeq \boldsymbol{0}$ denotes the covariance matrix of the AN. Therefore, the received signal at Bobs can be expressed as 
\begin{equation}\label{receive}
    \boldsymbol{y_{B}} = \boldsymbol{H}_{B}\boldsymbol{x}+\boldsymbol{z_{B}},
\end{equation}
where $\boldsymbol{z_B}\in \mathbb{C}^{I\times 1}$ denotes the additive white Gaussian noise (AWGN) with variance $\sigma_{B}^2$.
% We denote as $\bar{\boldsymbol{x}}\in \mathbb{C}^{L\times 1}$ contains the vector $\boldsymbol{x}$ $M$ times.

Similarly, the wiretap channel between all PAs and eavesdroppers can be expressed as
\begin{equation}
\boldsymbol{H}_{E} = \boldsymbol{H}_E^{(2)} \boldsymbol{H}^{(1)} \in \mathbb{C}^{K\times N} ,\,
\end{equation}
where $\boldsymbol{H}_E^{(2)}\in \mathbb{C}^{K\times L}$ denotes the free-space propagation from each PA to the $K$ eavesdroppers, with entries
\begin{equation} \label{channel_eve}
    \left( \boldsymbol{H}_E^{(2)}\right)_{k,(n,m)}=h^{(2)}_{e_k,n,m}=\frac{\eta^{\frac{1}{2}}e^{-j\frac{2\pi}{\lambda}\lVert \boldsymbol{\psi}_{E_k}-\tilde{\boldsymbol{\psi}}_{n,m} \rVert }}{\lVert \boldsymbol{\psi}_{E_k}-\tilde{\boldsymbol{\psi}}_{n,m} \rVert},
\end{equation}
where $\boldsymbol{\psi}_{E_k}$ denotes the position of the $k$-th Eve. The per-waveguide channel between the $n$-th waveguide and the $k$-th Eve is then obtained as
\begin{equation}
    \left( \boldsymbol{H}_E\right)_{k,n}=\displaystyle\sum_{\substack{m=1}}^{M} h^{(2)}_{e_k,n,m} h^{(1)}_{n,m}
\end{equation}
and the per-waveguide channel vector for Eve $k$ is $\boldsymbol{h}_{E,k}=[\left( \boldsymbol{H}_E\right)_{k,1}, ..., \left( \boldsymbol{H}_E\right)_{k,N}]^T\in \mathbb{C}^{N\times 1}$. It should be highlighted that \eqref{receive} also applies to Eves, whose AWGN variance is $\sigma_E^2$.

Therefore, the data rate of the $i$-th Bob and the data rate at the $k$-th Eve for the $i$-th Bob's data stream are given, respectively, as
{\small
\begin{equation}
\begin{split}
   R_\mathrm{B,i} =&  \log_2\Bigg(1+\frac{\lVert \boldsymbol{h}_{B,i}^H\,{\boldsymbol{w}}_i\rVert^2}
    {\displaystyle\sum_{\substack{m=1\\m\neq i}}^{I} \bigl|\mathbf{h}_{B,i}^H\boldsymbol{w}_m\bigr|^2 + \lVert \boldsymbol{h}_{B,i}^H\,\boldsymbol{R}_m\,\boldsymbol{h}_{B,i}\rVert + \sigma_B^2}\Bigg),\\
   R_\mathrm{E,{k,i}} =&  \log_2\Bigg(1+\frac{\lVert \boldsymbol{h}_{E,k}^H\,{\boldsymbol{w}}_i\rVert^2}
    {\displaystyle\sum_{\substack{m=1\\m\neq i}}^{I} \bigl|\mathbf{h}_{E,k}^H\boldsymbol{w}_m\bigr|^2 + \lVert \boldsymbol{h}_{E,k}^H\,\boldsymbol{R}_m\,\boldsymbol{h}_{E,k}\rVert + \sigma_E^2}\Bigg),
\end{split}
\end{equation}
}
which can be equivalently written as
{\small
\begin{equation}
\begin{split}\label{eq:R_alt}
   R_\mathrm{B,i} =&  \log_2\Bigg(1 \hspace{-0.75mm}+\hspace{-0.75mm}\frac{\operatorname{Tr}\left(\boldsymbol{H}_{B,i}\,{\boldsymbol{W}}_i\right)}
    {\displaystyle\sum_{\substack{m=1\\m\neq i}}^{I} \operatorname{Tr}\left(\boldsymbol{H}_{B,i}\,{\boldsymbol{W}}_m\right) + \operatorname{Tr}\left(\boldsymbol{H}_{B,i}\,{\boldsymbol{R}}_m\right) + \sigma_B^2}\Bigg),\\
   R_\mathrm{E,{k,i}} =&  \log_2\Bigg(1\hspace{-0.75mm}+\hspace{-0.75mm}\frac{\operatorname{Tr}\left(\boldsymbol{H}_{E,k}\,{\boldsymbol{W}}_i\right)}
    {\displaystyle\sum_{\substack{m=1\\m\neq i}}^{I}\operatorname{Tr}\left(\boldsymbol{H}_{E,k}\, {\boldsymbol{W}}_m \right)+ \operatorname{Tr}\left(\boldsymbol{H}_{E,k}\,{\boldsymbol{R}}_m\right) + \sigma_E^2}\Bigg),
\end{split}
\end{equation}
}where $\operatorname{Tr}(\cdot)$ denotes the trace operator, while $\boldsymbol{H}_{B,i} = \boldsymbol{h}_{B,i} \boldsymbol{h}_{B,i}^H$, $\boldsymbol{H}_{E,k} = \boldsymbol{h}_{E,k} \boldsymbol{h}_{E,k}^H$ and ${\boldsymbol{W}}_i={\boldsymbol{w}}_i {\boldsymbol{w}}_i^H$. Consequently, the SR is given as
\begin{equation}\label{eq:SR}
    \mathrm{SR} = \displaystyle\min_{{i,k}}[R_\mathrm{B,i}-R_\mathrm{E,k,i}]^+.
\end{equation}

\section{Problem Formulation with perfect CSI}\label{sec3}

First, we assume that the eavesdroppers pretend to be legitimate users and transmit uplink pilot signals. Consequently, the BS can obtain the CSI for both Bob and Eve \cite{huawai, optimization2}. Using this prior knowledge, the BS can enhance the signal quality for Bob while effectively mitigating the information leakage to Eve. In this section, we assume that perfect CSI is available.
Furthermore, as indicated by the SR expression, the additional spatial DoFs provided by the PAs allow not only the design of the baseband beamformer and the AN matrix, but also the reconfiguration of the channels. This flexibility allows the joint optimization of $\boldsymbol{W},\boldsymbol{R}_m$, and the PA position matrix $\boldsymbol{\tilde{x}}_{P}\in\mathbb{R}^{N\times M}$, which contains the positions of all PAs. Thus, from \eqref{eq:R_alt}, \eqref{eq:SR}, the optimization problem can be formulated as 
\begin{equation*} \tag{\textbf{P1}}\label{eq:basic_opt}
    \begin{array}{cl}
    \mathop{\max}\limits_{\boldsymbol{W}, \boldsymbol{R}_m, \boldsymbol{\tilde{x}}_P} &\mathrm{SR} \\
        \textbf{s.t.} & \mathrm{C}_1: \, \tilde{x}_{n,m}\in [0, D],\quad \forall\, n\in\mathcal{N}, \forall\, m\in \mathcal{M},  \\
        & \mathrm{C}_2: \, \tilde{x}_{n,m+1}\,-\,\tilde{x}_{n,m}\geq\Delta, \\
        &\quad\quad \forall\, n\in\mathcal{N},\, \forall m \in \mathcal{M}-\{M\},\\
        & \mathrm{C}_3: \operatorname{Tr}\Bigl(\displaystyle\sum_{\substack{i=1}}^{I}\boldsymbol{W}_i + \boldsymbol{R}_m\Bigr) \leq P, \\
        & \mathrm{C}_4: \boldsymbol{W}_i\succeq 0,\quad \boldsymbol{R}_m\succeq 0, \quad\forall\, i\in\mathcal{I},  
    \end{array}
\end{equation*}
where $P$ denotes the total transmit power available at the BS. Constraint $\mathrm{C}_1$ ensures that the optimized location of the PA remains within the physical boundaries of the waveguide, while constraint $\mathrm{C}_2$ guarantee that the antenna spacings should be no smaller than the minimum distance $\Delta$ to avoid the antenna coupling. Constraint $\mathrm{C}_3$ guarantees that the total transmit power does not exceed the available power budget. Finally, constraint $\mathrm{C}_4$ requires that both $\boldsymbol{W}$ and $\boldsymbol{R}_m$ are positive semidefinite matrices.

\subsection{AN-aided Secure Beamforming with a Single Waveguide}
In this section, we examine a special case of problem~\eqref{eq:basic_opt} in which the system consists of a single waveguide with just one PA mounted on it, i.e., \(N = 1,\, M=1\). In this scenario, both \(\boldsymbol{w}\), \(\boldsymbol{R}_m\) and $\boldsymbol{\tilde{x}}_P$ reduce to scalars and constraint $\mathrm{C}_2$ of problem \eqref{eq:basic_opt} can be omitted, yielding a more tractable formulation. The resulting optimization problem can be expressed as
\begin{equation*} \tag{\textbf{P2}}\label{eq:opt_single}
    \begin{array}{cl}
    \mathop{\max}\limits_{w, R_m, \tilde{x}_P} &\mathrm{SR} \\
        \text{\textbf{s.t.}}& \mathrm{C}_1: \, \tilde{x}_P\in [0, D], \\
        & \mathrm{C}_2: \lVert w \rVert^2+R_m \leq P. \\
    \end{array}
\end{equation*}
We note that the objective in~\eqref{eq:opt_single} remains non-convex. To address this challenge, we propose an alternating optimization method that decomposes the original problem into two convex subproblems which are solved iteratively until convergence. They are formulated as 
\begin{equation*} \tag{\textbf{P2.1}}\label{eq:opt_single_x}
    \begin{array}{cl}
    \mathop{\max}\limits_{\tilde{x}_P} &\mathrm{SR}  \\
        \text{\textbf{s.t.}}& \mathrm{C}_1: \, \tilde{x}_P\in [0, D], \\
    \end{array}
\end{equation*}
where $w$ and $R_m$ are considered fixed, and
\begin{equation*} \tag{\textbf{P2.2}}\label{eq:opt_single_w}
    \begin{array}{cl}
    \mathop{\max}\limits_{w, R_m} &\mathrm{SR} \\
        \text{\textbf{s.t.}}& \mathrm{C}_2: \, \lVert w \rVert^2+R_m \leq P, \\
    \end{array}
\end{equation*}
where $\tilde{x}_P$ is now fixed.

\begin{lemma}\label{l1}
The objective function in optimization problem~\eqref{eq:opt_single_x} is univariate, and a closed-form solution for $\tilde{x}_P$ is derived for given values of $w$ and $R_m$.
\end{lemma}
\begin{IEEEproof}
The proof is presented in Appendix A.
\end{IEEEproof}

Having determined the optimal $\tilde{x}_P$ from~\eqref{eq:opt_single_x}, we now address problem~\eqref{eq:opt_single_w}. Specifically, we first solve constraint $\mathrm{C}_2$ for $R_m$ and by substituting this expression into equation~\eqref{eq:SR} and performing some algebraic manipulations, the objective function can be written as
\begin{equation}\label{eq:SR_fixed_x}
    \mathrm{SR} = \log_2{\left( \frac{(\eta P+r_B^2 \sigma_B^2)(\eta R_m+r_E^2 \sigma_E^2)}{(\eta P+r_E^2 \sigma_E^2)(\eta R_m+r_B^2 \sigma_B^2)}  \right)},
\end{equation}
where $r_B = \lVert \boldsymbol{\psi}_{B}-\tilde{\boldsymbol{\psi}}_{P} \rVert$ and $r_E = \lVert \boldsymbol{\psi}_{E}-\tilde{\boldsymbol{\psi}}_{P} \rVert$. 
\begin{lemma}\label{l2}
The objective function defined in \eqref{eq:SR_fixed_x} is monotone with respect to $R_m$. Specifically, if $r_B^2 \sigma_B^2 < r_E^2 \sigma_E^2$, the objective function is monotonically decreasing, and thus its maximum occurs at $R_m = R_m^{\mathrm{min}} = 0$. Conversely, if $r_B^2 \sigma_B^2 > r_E^2 \sigma_E^2$, it is monotonically increasing and has its maximum at $R_m = R_m^{\mathrm{max}} = P$.
\end{lemma} 
\begin{IEEEproof} 
The proof is given in Appendix B.
\end{IEEEproof} 
\begin{remark} 
In the proof of Lemma \ref{l1} for the case where the AWGN variances are equal (i.e., $\sigma_B^2 = \sigma_E^2$), a common assumption in the literature, it follows that the optimal strategy is to allocate the entire power budget to Bob without injecting any AN, provided that Bob is physically closer to the transmitting antenna than Eve, since the condition simplifies to $r_B<r_E$. This condition is generally achievable through the reconfigurability of PAs. However, when operating with a single waveguide, it is not always feasible to ensure a PA position that places Bob closer to the transmit source than Eve. Consequently, the use of additional waveguides becomes imperative to ensure the desired proximity advantage for secure communications.
\end{remark}

\subsection{AN-aided Secure Beamforming with Multiple Waveguides}

\begin{figure}
    \centering
    % \includegraphics[height=5.3cm]
    % \vspace{-4mm}
    \includegraphics[width=1\columnwidth]{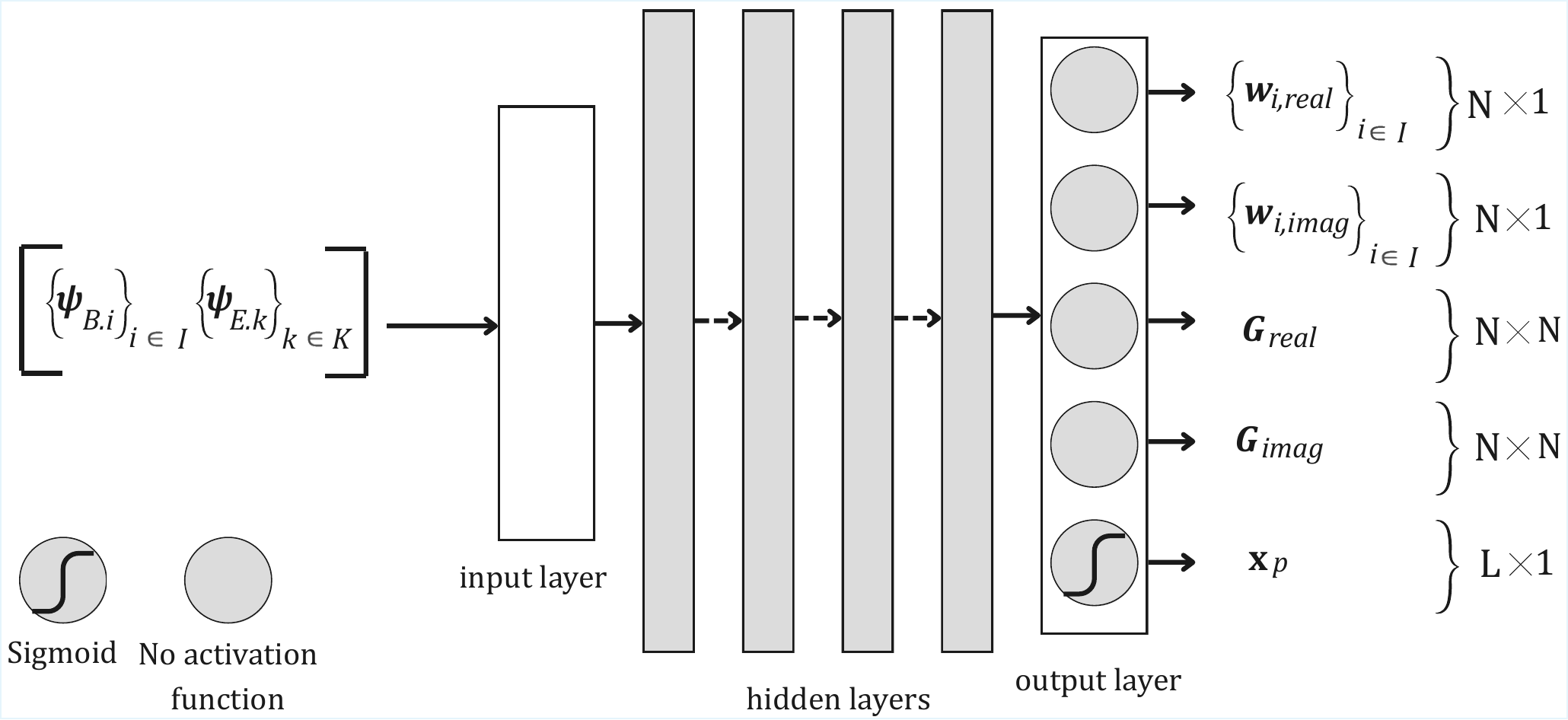}
    \vspace{-2mm}
    \caption{The DNN architecture.}
    \vspace{-4mm}
    \label{fig:DNN_arch}
\end{figure}

Observing \eqref{eq:basic_opt}, we identify two major challenges. First, the objective function is inherently non-convex due to the fractional form and the complicated interdependencies between the optimization variables, making it difficult to find a global optimum. Second, the tight coupling of the optimization variables further increases the complexity of the problem. To address these difficulties, we propose a DNN-aided framework which can be used to obtain a near-optimal solution to highly non-convex problems \cite{bouzinis, bozanis_2} such as problem \eqref{eq:basic_opt}. DNN-aided optimization has been proved to provide optimal solutions to optimization problems under the mild assumption of the time-sharing condition \cite{DNN_JSAC}, thus it is a promising alternative for non-convex optimization problems. The proposed DNN’s architecture is given in Fig. \ref{fig:DNN_arch}. Given the locations $\boldsymbol{\psi}_{B}$ of Bobs and $\boldsymbol{\psi}_{E}$ of Eves in the room, the DNN is trained in an online fashion to estimate the optimization variables. It is clarified that, unlike offline approaches, where multiple samples are used for training and afterwards the testing stage follows, online learning has no testing stage since for each new generated sample a new dedicated network is trained, thus the generalization issue does not exist. As such, online learning is a great tool to tackle such complex non-convex optimization forms. 

The beamforming vector $\boldsymbol{w}_i\in \mathbb{C}^{N\times 1},$  for each legitimate user Bob $i$, is emitted in two blocks of $N$ real numbers with no activation, interpreting the first as $\Re\{\boldsymbol{w}_i\}$ and the second as $\Im\{\boldsymbol{w}_i\}$. Recombining $w_i = \Re\{w_{i}\}+i\,\Im\!\{w_{i}\}$ allows the network to utilize the full complex beamforming space directly. Once we acquire each beamforming vector $\boldsymbol{w}_i$ we form the rank-1 covariance matrix $\boldsymbol{W}_i$ that automatically ensures $\boldsymbol{W}_i\succeq 0$. Next, the DNN  outputs two blocks, each of size $N\times N$  with no activation, which are interpreted as the real and imaginary parts of an unconstrained complex matrix $\boldsymbol{G}\in \mathbb{C}^{N\times N}$. Concretely, if $\boldsymbol{G}_{\mathrm{real}}$ is the first $N^2$ output reshaped to $N\times N$ and $\boldsymbol{G}_{\mathrm{imag}}$ is the next $N^2$ output reshaped to $N\times N$, then $\boldsymbol{G}=\boldsymbol{G}_{\mathrm{real}}+i\,\boldsymbol{G}_{\mathrm{imag}}$. Hence, the AN covariance matrix is formed as $\boldsymbol{R}_m=\boldsymbol{G}\boldsymbol{G}^H$, which by construction is Hermitian positive semidefinite. Furthermore, the PA position matrix $\boldsymbol{\tilde{x}}_P\in \mathbb{R}^{M\times N}$ is produced by first taking the corresponding slice of the final linear output $\boldsymbol{\tilde{x}}_{P, f}\in \mathbb{R}^{1\times MN}$ and applying a sigmoid activation to each element, forcing it into the $[0, 1]$ interval. We then reshape to $(M, N)$ and multiply by the room length $D$. This construction automatically guarantees that $0\leq \tilde{x}_{m,n}\leq D,\, \forall\, n\in\mathcal{N}, \forall\, m\in \mathcal{M}$, ensuring that every PA stays within the allowed boundaries.

For the DNN training, a carefully designed loss function is essential, as it encodes the optimization objective and any task-specific constraints. In our case, the loss function is defined as
\begin{equation}\label{eq:loss1}
    \mathrm{Loss_1} = -\mathrm{SR}+ U_1 A_{\mathrm{pen}_1} + U_2 A_{\mathrm{pen}_2},
\end{equation}
where the minus sign was introduced for the DNN to maximize the SR of the PAS, while $U_i>0$ and $A_{\mathrm{pen}_i},\,i\in [1,2]$ are penalty terms to ensure that the constraints hold. 
Term $A_{\mathrm{pen}_1}$ represents the square of the constraint violation. However, for constraint $\mathrm{C}_3$ this will drive the DNN’s solution to the region where $\operatorname{Tr}\Bigl(\displaystyle\sum_{\substack{i=1}}^{I}\boldsymbol{W}_i + \boldsymbol{R}_m\Bigr) = P$, which is not necessarily the optimal solution since it can cause increased interference between users. Similar insights hold for the term $A_{\mathrm{pen}_2}$ with constraint $\mathrm{C}_2$. Our goal is to discourage violations and not force equality, thus we adopt a one-sided penalty that remains inactive when the constraint is satisfied. For that reason, we use the activation function, \text{ReLU}, which is defined as the non-negative part of its argument, i.e., the ramp function, given as
\begin{equation}
    \text{ReLU}(x) = \max(0,x) = \dfrac{x+|x|}{2} = \begin{cases}
  x, & x>0\\
  0, & x\le 0,
\end{cases}
\end{equation}
which is differentiable everywhere, except at zero. Then, we define the penalty terms as follows:
\begin{equation}
    A_{\mathrm{pen}_1} = \text{ReLU}\left(\operatorname{Tr}\Bigl(\displaystyle\sum_{\substack{i=1}}^{I}\boldsymbol{W}_i + \boldsymbol{R}_m\Bigr) - P\right)^2
\end{equation}
\begin{equation}
    A_{\mathrm{pen}_2} \!=\! \dfrac{1}{(N-1)M}\sum_{m=1}^{M}\sum_{n=2}^{N}
                \bigl[\mathrm{ReLU}\bigl(\Delta -(\tilde{x}_{n,m}-\tilde{x}_{n-1,m})\bigr)\bigr]^{2} \!.
\end{equation}

\section{Problem Formulation With Imperfect CSI}\label{sec4}

In practical systems, many eavesdroppers are passive and non-cooperative, so the transmitter cannot acquire precise CSI or location, and it operates instead with a noisy estimate of Eve’s effective downlink response. Therefore, the BS observes a complex vector $\hat{\mathbf{h}}_{E,k}\in \mathbb{C}^{N\times 1}$ that incorporates the phases and path losses of $N$ waveguide links, each link being the aggregated contribution of the $M$ PAs of that waveguide. However, any residual error, caused by noisy pilots, insufficient training, hardware offsets, and uncertainty in Eve's location estimation, is added up as a small disturbance directly to the same $N$-dimensional channel vector. Thus, this residual is represented as an additive perturbation as
\begin{equation}
    {\mathbf{h}}_{E,k} = \hat{\mathbf{h}}_{E,k} + \boldsymbol{\Delta h}_k, 
\end{equation}
where $\boldsymbol{\Delta h}_k\in \mathbb{C}^{N\times 1}$ denotes the channel error for $k$-th Eve and is bounded by a quadratic form as
\begin{equation}
    \boldsymbol{\Delta h}_k^H \;\boldsymbol{\Phi}_k\;\boldsymbol{\Delta h}_k \;\le\; 1,
\end{equation}
where $\boldsymbol{\Phi}_k$ is a Hermitian positive definite matrix and defines the ellipsoidal uncertainty in the channel estimation \cite{boyd_2}. Ellipsoidal uncertainty sets are well suited for wireless channels because they capture anisotropy of the estimation error, they lead to tractable robust constraints, and they reduce to a sphere when appropriate, thus avoiding conservatism. From \eqref{channel_eve}, the channel is a function of $k$-th Eve’s physical coordinates $\boldsymbol{p}_k=(x_k,y_k,z_k)$, which can also be expressed as $\boldsymbol{p}_k=\boldsymbol{\hat{p}}_k+\mathbf{\Delta p}_k$, where $\boldsymbol{\hat{p}}$ is the estimation of Eve's position, while $\mathbf{\Delta p}_k$ the position offset. Thus, the new channel of $k$-th Eve can be approximated with a first-order Taylor expansion as
\begin{equation}
    \mathbf{h}_{E,k}(\mathbf{\hat{p}}_k+\mathbf{\Delta p}_k)\approx \mathbf{\hat h}_{E,k}(\hat{\mathbf{p}_k}) \;+\; \mathbf{J}_k\mathbf{\Delta p}_k ,
\end{equation}
where $\mathbf{\hat h}_{E,k}(\hat{\mathbf{p}_k})$ is the estimated channel vector and 
\begin{equation}
    \mathbf{J}_k \;\triangleq\; 
\left.\frac{\partial\, \mathbf{h}_{E,k}(\mathbf{p})}{\partial \mathbf{p}}\right|_{\mathbf{p}=\hat{\mathbf{p}}_k}
\in \mathbb{C}^{N\times 3}
\end{equation}
is the Jacobian that measures how each channel is affected with shifts in $x$, $y$, or $z$ coordinate. Assuming that the position error is zero-mean with diagonal covariance, $\boldsymbol{\Sigma}_{xyz}=\text{diag}(\sigma_x^2, \sigma_y^2, \sigma_z^2)$ lies inside the set $\{\mathbf{\Delta p}_k: \mathbf{\Delta p}_k^T \boldsymbol{\Sigma}_{xyz}^{-1} \mathbf{\Delta p}_k \leq 1\}$. Therefore, each offset is mapped through the Jacobian, $\mathbf{\Delta h}_k= \mathbf{J}_k \mathbf{\Delta p}_k$, and every vector is rescaled and rotated, thus the original error region in the position domain is transformed into an ellipsoid in the channel domain. The ellipsoidal uncertainty matrix after some mathematical manipulations becomes $\boldsymbol{\Phi}_k=\mathbf{J}_k\boldsymbol{\Sigma}_{xyz}\mathbf{J}_k^H$. Since the derivative entries in $\mathbf{J}_k$ are numerically large for the PAs closest to Eves and small for distant ones, $\mathbf{\Phi}_k$ imposes tighter tolerances where the channel is more sensitive to user motion and looser tolerances where it is not. Thus, the single isotropic ball of location offsets is converted into a directionally weighted ellipsoid in the $N$-dimensional channel space, providing an accurate description of CSI uncertainty. 

The SINR expression for imperfect CSI for the $k$-th Eve when the $i$-th Bob is served is given as
\begin{equation}
    \gamma_{E_{k,i}}(\mathbf{\Delta h}_k) \hspace{-0.75mm}= \hspace{-0.75mm}\dfrac{\lVert(\hat{\mathbf{h}}_{E_k}+\mathbf{\Delta h}_k)^H \boldsymbol{w}_i\rVert ^2}{(\hat{\mathbf{h}}_{E_k}+\mathbf{\Delta h}_k)^H \mathbf{Q}_{k,i} (\hat{\mathbf{h}}_{E_k}+\mathbf{\Delta h}_k) + \sigma_E^2},
\end{equation}
where $\mathbf{Q}_{k,i} = \mathbf{R}_m + \sum_{\substack{m\neq i}} \mathbf{W}_m$ is the interference from other users summed with the AN covariance. Since the perturbation $\boldsymbol{\Delta h}_k$ resides in the ellipsoid,
\begin{equation}\label{eq:ellipsoid}
\mathcal{H}_k = \{\boldsymbol{\Delta h}_k: \boldsymbol{\Delta h}_k^H \;\boldsymbol{\Phi}_k^{-1}\;\boldsymbol{\Delta h}_k \;\le\; 1\},
\end{equation}
to guarantee security, we upper-bound Eve’s SINR over the entire uncertainty set with a design scalar $\lambda_{k,i}>0$. Consequently, we get
\begin{equation}\label{eq:worst_case}
\begin{aligned}
    &\displaystyle\max_{\boldsymbol{\Delta h}_k\in \mathcal{H}_k} \gamma_{E_{k,i}}(\mathbf{\Delta h}_k)\leq \dfrac{\lVert (\hat{\mathbf{h}}_{E_k}+\mathbf{\Delta h}_k)^H \boldsymbol{w}_i\rVert ^2}{\lambda_{k,i}} ={:}\gamma_{E_{k,i}}^{wc},
\end{aligned}
\end{equation}
where $\gamma_{E_{k,i}}^{wc}$ denotes the highest SINR that the $k$-th Eve can experience when the $i$-th Bob is served. The above condition is satisfied when
\begin{equation}
    \displaystyle\min_{\boldsymbol{\Delta h}_k\in \mathcal{H}_k} (\hat{\mathbf{h}}_{E_k}+\mathbf{\Delta h}_k)^H \mathbf{Q}_{k,i} (\hat{\mathbf{h}}_{E_k}+\mathbf{\Delta h}_k) + \sigma_E^2\geq \lambda_{k,i}.
\end{equation}

We now set 
\begin{equation} \label{fx}
    f(\mathbf{\Delta h}_k) = (\hat{\mathbf{h}}_{E_k}+\mathbf{\Delta h}_k)^H \mathbf{Q}_{k,i}   (\hat{\mathbf{h}}_{E_k}+\mathbf{\Delta h}_k) + \sigma_E^2 - \lambda_{k,i} ,
\end{equation}
which could be rewritten as
\begin{equation}
\begin{aligned}\label{eq:fx2}
    f(\mathbf{\Delta h}_k) &= \mathbf{\Delta h}_k^H \mathbf{Q}_{k,i} \mathbf{\Delta h}_k + 2\Re\{\hat{\mathbf{h}}_{E_k}^H \mathbf{Q}_{k,i} \mathbf{\Delta h}_k\} \\&+ (\hat{\mathbf{h}}_{E_k}^H \mathbf{Q}_{k,i} \hat{\mathbf{h}}_{E_k} + \sigma_E^2-\lambda_{k,i}),
\end{aligned}
\end{equation}
which is also a quadratic function.
Moreover, the perturbation vector $\boldsymbol{\Delta h}_k$ is confined to the ellipsoid
\begin{equation}\label{gx}
    g(\mathbf{\Delta h}_k) = \boldsymbol{\Delta h}_k^H \;\boldsymbol{\Phi}_k^{-1}\;\boldsymbol{\Delta h}_k-1\leq0.
\end{equation}

Maximizing Eve's SINR directly over the infinite set $\mathcal{H}_k$ would render the optimization intractable. Instead, since both $f(\cdot)$ and $g(\cdot)$ are quadratic with respect to $\mathbf{\Delta h}_k$, the S-procedure can be used to replace the semi-infinite constraint by a single linear matrix inequality (LMI) \cite{LMI_1, boyd}. Thus,
\begin{comment}
\begin{lemma}\label{l3}
Let a function $f_i(\mathbf{x})$, $i\in \{1, 2\}$, $\mathbf{x}\in \mathbb{C}^{J\times 1}$, which is defined as
\[
f_i(\mathbf x)
=\mathbf x^H A_i\,\mathbf x \;+\;2\,\Re\{\,\mathbf b_i^H\mathbf x\}\;+\;c_i,
\quad i\in\{1,2\},
\]
where $\mathbf{A}_i\in \mathbb{C}^{J\times J}$, $\mathbf{b}_i\in \mathbb{C}^{J\times 1}$ and $c_i\in \mathbb{R}$. 
Assume there exists some \(\mathbf x_0\) such that 
\(\;f_2(\mathbf x_0)<0\;\). Then, the implication
\[
f_2(\mathbf x)\le0
\;\Longleftrightarrow\;
f_1(\mathbf x)\ge0
\]
holds if and only if there is a scalar \(\tau\ge0\) for which
\[
\begin{pmatrix}
A_1 & b_1\\[4pt]
b_1^H & c_1
\end{pmatrix}
\;-\;
\tau\,
\begin{pmatrix}
A_2 & b_2\\[4pt]
b_2^H & c_2
\end{pmatrix}
\;\succeq\;0.
\]
\end{lemma}
\end{comment}
a non-negative multiplier $\tau_{k,i}\ge0$ is obtained such that the LMI becomes
\begin{equation}\label{eq:lmi}
\begin{aligned}
    &\boldsymbol{M}_{{k,i}}(\tau_{k,i}, \lambda_{k,i}, \boldsymbol W,\;\boldsymbol R_m,\;\tilde{\boldsymbol x}_P) = \\
    &\begin{bmatrix}
    \mathbf{Q}_{k,i} - \tau_{k,i}\,\mathbf{\Phi}_k^{-1} & \mathbf{Q}_{k,i}\,\hat{ \mathbf{h}}_{E,k} \\[6pt]
    \hat{\mathbf{h}}_{E,k}^H\,\mathbf{Q}_{k,i} & \hat{\mathbf{h}}_{E,k}^H\,\mathbf{Q}_{k,i}\,\hat{\mathbf{h}}_{E,k}
   +\sigma_E^2-\lambda_{k,i}+\tau_{k,i}
\end{bmatrix}
\succeq 0.
\end{aligned}
\end{equation}
This positive semidefinite condition ensures that the worst-case SINR of the $k$-th Eve when the $i$-th Bob is served obeys \eqref{eq:ellipsoid} and \eqref{eq:worst_case}.

Therefore, since the worst-case $\gamma_{E_{k,i}}(\mathbf{\Delta h}_k)$ is upper-bounded, the optimization problem can be reformulated as
\begin{equation*} \tag{\textbf{P3}}\label{eq:final_opt}
    \begin{array}{cl}
        \displaystyle\mathop{\max}_{\substack{
        \boldsymbol W,\;\boldsymbol R_m,\;\tilde{\boldsymbol x}_P,\\
        \lambda_{k,i},\;\tau_{k,i}
        }} &\widehat{\mathrm{SR}}=\mathrm{\displaystyle\min_{k,i} \left[R_{B,i}-\log_2\left(1+\gamma_{E_{k,i}}^{wc}\right)\right]} \\
        \textbf{s.t.} & \mathrm{C}_1: \, \tilde{x}_{n,m}\in [0, D],\quad \forall\, n\in\mathcal{N}, \forall\, m\in \mathcal{M}_n , \\
        & \mathrm{C}_2: \, \tilde{x}_{n,m+1}\,-\,\tilde{x}_{n,m}\geq\Delta, \\
        &\quad\quad \forall\, n\in\mathcal{N},\, m=1,\forall m \in \mathcal{M}-\{M\},\\
        & \mathrm{C}_3: \operatorname{Tr}\Bigl(\displaystyle\sum_{\substack{i=1}}^{I}\boldsymbol{W}_i + \boldsymbol{R}_m\Bigr) \leq P, \\
        & \mathrm{C}_4: \boldsymbol{M}_{{k,i}}(\tau_{k,i}, \lambda_{k,i}, \boldsymbol W,\;\boldsymbol R_m,\;\tilde{\boldsymbol x}_P)\succeq 0, \\
        &\quad \quad\forall i\in \mathcal{I},\,\forall k\in \mathcal{K}, \\
        & \mathrm{C}_5: \boldsymbol{W}_i\succeq 0,\quad \boldsymbol{R}_m\succeq 0,\\
        & \quad \quad\tau_{k,i}\ge0,\quad \lambda_{k,i}>0,\quad\forall\, i\in\mathcal{I},\, \forall k\in \mathcal{K}, 
    \end{array}
\end{equation*}
where $C_4$ is the constraint related to the estimation errors in the eavesdroppers' channels.
The problem \eqref{eq:final_opt} is still non-convex because the variables that enter the objective and $\mathrm{C}_4$ are highly coupled. 

Although the S-procedure does not, in general, convexify the problem, the induced LMI acts as a verifiable sufficient condition for robust feasibility. This is because the SINR lower bound holds for every channel vector in the ellipsoidal uncertainty set if its minimum eigenvalue is non-negative. During training, we enforce this condition by adding a smooth penalty on the negative part of the smallest eigenvalue, which preserves end-to-end differentiability and allows optimization via a DNN. In effect, the S-procedure replaces an intractable semi-infinite constraint with a finite and differentiable condition, making the DNN-based design both practical and reliable.

The main DNN architecture of Section \ref{sec3} is retained, however two additional output blocks, each of size $(I\times K)$, produce the non-negative variables $\tau_{k,i}$ and the positive variables $\lambda_{k,i}$ via a soft-plus activation. Hence, the loss function for \eqref{eq:final_opt} is given by
\begin{equation}
    \mathrm{Loss}_2 = -\widehat{\mathrm{SR}} + U_1 A_{\mathrm{pen}_1} + U_2 A_{\mathrm{pen}_2} + U_3 A_{\mathrm{pen}_3},
\end{equation}
where $A_{\mathrm{pen}_1}$ and $A_{\mathrm{pen}_2}$ that enforce power budget and minimum spacing constraints are the same as in \eqref{eq:loss1}. The extra term $A_{\mathrm{pen}_3}$ is defined as
\begin{equation}
   A_{\mathrm{pen}_3} = \left\langle
  \mathrm{ReLU}\bigl(-\lambda_{\min}(\boldsymbol{M}_{_{k,i}})\bigr)
\right\rangle_{k,i} ,
\end{equation}
where $\lambda_{\min}(\boldsymbol{M}_{k,i})$ is the smallest eigenvalue of \eqref{eq:lmi}. This constraint drives the smallest eigenvalue of every matrix in \eqref{eq:lmi} to non-negative values, to ensure that the constraint holds.

\subsection{Complexity analysis} Let $H$ denote the typical width of a hidden layer in the DNN. In the perfect CSI scenario, the per‑iteration complexity scales as $\mathcal{O}\left(H^2+HN^2+N^3+I^2N^2+IN^2+IKN^2\right)$. However, in the imperfect CSI setting where the ellipsoidal uncertainty is tackled via the S‑procedure, an additional cubic term appears due to the minimum‑eigenvalue evaluation of a $(2N+1)\times (2N+1)$ LMI, yielding a complexity of $\mathcal{O}\left(H^2+HN^2+N^3+I^2N^2+IN^2+IKN^2+IKN^3\right)$.

\section{Numerical Results}\label{sec:Num}   

In this section, we numerically evaluate the performance of the proposed scheme by examining the mean achievable SR with respect to different variables and the CDF of the SRs through Monte Carlo simulations. As a point of reference, we consider a PAS architecture without AN \cite{huawai}. The simulation parameters are set as follows, unless stated otherwise. All dielectric waveguides are placed at a height of \( d = 2 \) m, with lengths equal to the side dimension \( D = 5 \) m, the noise power at both Bob and Eve is set to \( \sigma_B^2 = \sigma_E^2 = -90 \) dBm, the carrier frequency is set to \( f_c = 28\) GHz, and the effective refractive index of each dielectric waveguide is set to \( n_{\mathrm{eff}} = 1.4 \). Furthermore, we consider having one Bob, one Eve, two waveguides with four PAs mounted on each waveguide. 

The DNN consists of four hidden layers with 256 nodes, each followed by a \text{ReLU} activation function, except the last layer, as shown in Fig.~\ref{fig:DNN_arch}. The Adam optimizer was used with an initial learning rate of 0.0001, which was multiplied by 0.2 every 200 epochs. The total number of epochs per Monte Carlo iteration was 1500, while 500 Monte Carlo scenarios were examined for each mean SR value. The penalty factor $U_1$ is initialized inversely to the power budget $(1,30,300,3000,9000,15000)$ for $P=(20,15,10,5,0,-5,-10)$ \text{dBm} and then inflated by 20\% every 10 epochs until it reaches at most 100 times its starting value, after which it remains fixed. The minimum spacing constraint is multiplied by a constant coefficient $U_2=100$, chosen so that its gradient is on the same order as that of the SR term. Finally, the S‑procedure robustness penalty is multiplied by $U_3 = (2, 5, 10)\times 10^5$ for $\sigma_x^2,\sigma_y^2\in (0.05, 0.15, 0.3)$, respectively, and it grows or shrinks linearly with the severity of the uncertainty, maintaining the LMI term’s influence on the same scale as the secrecy objective. For notation simplicity, we set $\sigma_x^2=\sigma_y^2=\sigma^2$ for the discussion of the following figures. %{\color{red} Maybe show the constraint violation??}

\begin{figure}
    \centering
    \begin{tikzpicture}
    \begin{axis}[
        width=0.83\linewidth,
        xlabel = {Epochs},
        ylabel = {Secrecy Rate (bps/Hz)},
        ymin = 0,
        ymax = 11,
        xmin = 1,
        xmax = 1500,
        ytick = {0,2,4,...,10},
        xtick = {0, 500, 1000, 1500},         
        % },
        scaled y ticks=false, % Prevents multiplication by powers of 10
        grid = both,
        minor grid style={gray!25},
        major grid style={gray!50},
        legend columns=1, 
    legend entries={
    Perfect CSI,
    \shortstack[l]{%
      Imperfect CSI
    }},
        legend cell align = {left},
        legend style={font=\tiny},
        legend style={at={(0,1)},anchor=north west},
        legend image post style={scale=0.7}, % Adjust the line size in the legend
        % legend pos = north west
        ]
        \addplot[
        blue,
        %mark = triangle,
        %mark repeat = 0,
        mark size = 2,
        % only marks,
        line width = 1pt,
        style = solid,
        ]
        table[x index=0, y index=1, col sep=space]{Figures3/convergence_perfect.dat};
        \addplot[
        red,
        mark repeat = 500,
        mark size = 2,
        % only marks,
        line width = 1pt,
        style = solid,
        ]
        table[x index=0, y index=1, col sep=space]{Figures3/convergence_imperfect_paper.dat};
    \end{axis}
\end{tikzpicture}
    \caption{SR convergence with $P=0$ dBm and $\sigma^2=0.05$.}
    \label{fig:epoch_SR}
\end{figure}
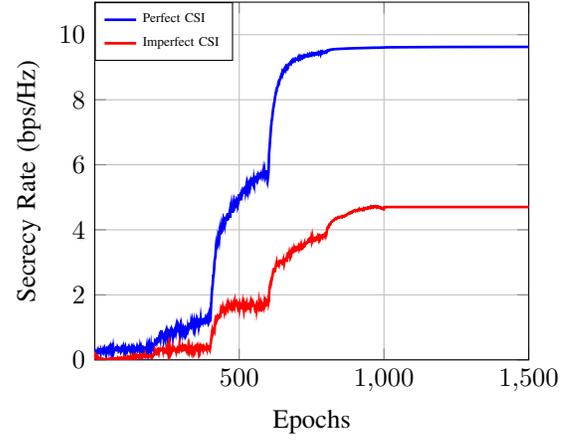

In Fig.~\ref{fig:epoch_SR}, the mean SR achieved by the proposed online DNN over 1500 training epochs is plotted for both perfect and imperfect CSI, for $P=0$ dBm, and for position error covariance $\sigma^2=0.05$. In the case of perfect CSI, the rate initially rises from zero, surging steeply for the first 750 epochs before settling around $9.5$ bps/Hz. Similar behavior is observed for imperfect CSI, however, the achievable rate is around $4.5$ bps/Hz. Both curves exhibit three broad ripples after each performance increase. These are triggered by the step learning‑rate decays. After each drop, the model briefly rebalances before resuming its ascent, while fluctuations reflect stochastic variability from random Bob and Eve placements. Overall, Fig.~\ref{fig:epoch_SR} shows that the perfect CSI run reaches its maximum slightly earlier than the imperfect CSI run, with the latter’s slower ascent and slightly lower achievable SR justified by the extra S‑procedure/LMI penalty that tightens the feasible set to achieve the best possible SR despite the uncertainty of Eves' location. Overall, faster convergence and higher SR are achieved when full CSI is available, and a tempered but still substantial SR is achieved when imperfect CSI is available. This quantifies the performance trade-off required to secure against imperfect knowledge of the eavesdroppers' channels.

\begin{figure} 
    \centering
    \begin{tikzpicture}
    \begin{axis}[
        width=0.83\linewidth,
        xlabel = {Side Length $D$ (m)},
        ylabel = {Secrecy Rate (bps/Hz)},
        ymin = 0,
        ymax = 12,
        xmin = 5,
        xmax = 30,
        xtick = {5,10,15,20,25,30},         
        % },
        scaled y ticks=false, % Prevents multiplication by powers of 10
        grid = both,
        minor grid style={gray!25},
        major grid style={gray!50},
        legend columns=1, 
        legend entries ={Perfect CSI, $\sigma^2=0.05$, $\sigma^2=0.15$, $\sigma^2=0.3$},
        legend cell align = {left},
        legend style={font=\tiny},
        legend style={at={(1,1)},anchor=north east},
        legend image post style={scale=0.7}, % Adjust the line size in the legend
        % legend pos = north west
        ]
        \addplot[
        green,
        mark repeat = 1,
        mark size = 2,
        % only marks,
        line width = 1pt,
        style = solid,
        ]
        table[x index=0, y index=1, col sep=space]{Figures3/SR_DLEN_1b_1e_2w_4pa_perfect.dat};
        \addplot[
        red,
        mark repeat = 1,
        mark size = 2,
        % only marks,
        line width = 1pt,
        style = solid,
        ]
        table[x index=0, y index=1, col sep=space]{Figures3/SR_DLEN_1b_1e_2w_4pa_imperfect.dat}; 
        \addplot[
        black,
        mark repeat = 1,
        mark size = 2,
        % only marks,
        line width = 1pt,
        style = solid,
        ]
        table[x index=0, y index=2, col sep=space]{Figures3/SR_DLEN_1b_1e_2w_4pa_imperfect.dat};
        \addplot[
        blue,
        mark repeat = 1,
        mark size = 2,
        % only marks,
        line width = 1pt,
        style = solid,
        ]
        table[x index=0, y index=3, col sep=space]{Figures3/SR_DLEN_1b_1e_2w_4pa_imperfect.dat};
    \end{axis}
\end{tikzpicture}
    \caption{SR vs. side length with $P=10$ dBm.}
    \label{fig:side_SR}
\end{figure}
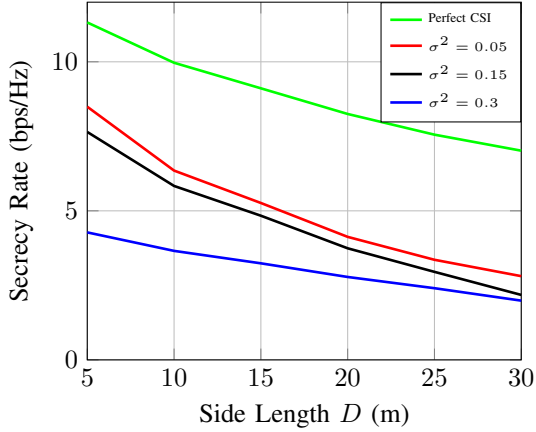

Fig.~\ref{fig:side_SR} depicts the mean SR as a function of the room side length $D$ at $P=10$ dBm, for perfect CSI and for three different covariances of Eves' position uncertainty, i.e., $\sigma^2=\{0.05, 0.15, 0.3\}$. Two consistent trends emerge. First, increasing the physical area decreases the SRs for all cases, since as $D$ increases, the Bob channels weaken, due to longer distances, limiting the spatial capabilities of the PAS. Second, the robustness to imperfect CSI differs significantly for different covariance values, reflecting the compounding effect of larger physical uncertainty ellipsoids and a larger feasible region in which the worst‑case Eve placements can align more closely with Bob’s effective beams. This forces the S‑procedure/LMI constraints to allocate more power to AN and less to the useful signal.

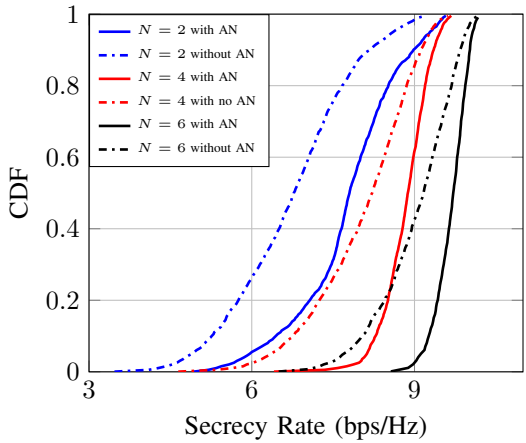
\begin{figure} 
    \centering
        \begin{tikzpicture}
        \begin{axis}[
            width=0.83\linewidth,
            ylabel = {CDF},
            xlabel = {Secrecy Rate (bps/Hz)},
            ymin = 0,
            ymax = 1,
            xmin = 3,
            xmax = 11,
            xtick = {3,6,9,12,15},         
            % },
            scaled y ticks=false, % Prevents multiplication by powers of 10
            grid = both,
            minor grid style={gray!25},
            major grid style={gray!50},
            legend columns=1, 
		legend entries ={$N=2$ with AN,$N=2$ without AN, $N=4$ with AN, $N=4$ with no AN, $N=6$ with AN, $N=6$ without AN},
            legend cell align = {left},
            legend style={font=\tiny},
            legend style={at={(0,1)},anchor=north west},
            legend image post style={scale=0.7}, % Adjust the line size in the legend
            % legend pos = north west
            ]
            \addplot[
            blue,
            mark repeat = 1,
            mark size = 2,
            % only marks,
            line width = 1pt,
            style = solid,
            ]
            table[x index=0, y index=1, col sep=space]{Figures/pin_2_CDF_rn.dat};
            \addplot[
            blue,
            mark repeat = 1,
            mark size = 2,
            % only marks,
            line width = 1pt,
            style = dashdotted,
            ]
            table[x index=0, y index=1, col sep=space]{Figures/pin_2_CDF_no_rn.dat};
            \addplot[
            red,
            mark repeat = 1,
            mark size = 2,
            % only marks,
            line width = 1pt,
            style = solid,
            ]
            table[x index=0, y index=1, col sep=space]{Figures/pin_4_CDF_rn.dat};
            \addplot[
            red,
            mark repeat = 1,
            mark size = 2,
            % only marks,
            line width = 1pt,
            style = dashdotted,
            ]
            table[x index=0, y index=1, col sep=space]{Figures/pin_4_CDF_no_rn.dat};
            \addplot[
            black,
            mark repeat = 1,
            mark size = 2,
            % only marks,
            line width = 1pt,
            style = solid,
            ]
            table[x index=0, y index=1, col sep=space]{Figures/pin_6_CDF_rn.dat}; 
            \addplot[
            black,
            mark repeat = 1,
            mark size = 2,
            % only marks,
            line width = 1pt,
            style = dashdotted,
            ]
            table[x index=0, y index=1, col sep=space]{Figures/pin_6_CDF_no_rn.dat};
        \end{axis}
    \end{tikzpicture}
    \caption{CDF of SR for various values of $N$ with $M=1$ and $P=10$ dBm.}
    \label{fig:CDF}
\end{figure}

Fig. \ref{fig:CDF} shows the CDF of the SR for \(P = 10\) dBm, comparing the proposed and the benchmark scheme over different values of \(N\) for $M=1$, aiming to study the importance of AN in the system design. We observe that incorporating AN into the PAS consistently outperforms the PAS benchmark throughout the CDF, highlighting the critical role of AN in enhancing system security and reliability, while both PASs do not experience any outage, illustrating the importance of the reconfigurability of PA positions. Nevertheless, the additional DoF provided by AN is critical for improving security in any configuration, delivering the largest gains in the lower tail of the SR CDF, where the Eves channels are stronger, while maintaining a notable gain at the median.
%notable gap at the median.     

\begin{figure} 
    \centering
    \begin{tikzpicture}
    \begin{axis}[
         width=0.83\linewidth,
          xlabel = {Transmit Power $P$ (dBm)},
          ylabel = {Secrecy Rate (bps/Hz)},
          ymin = 0, ymax = 17.5, xmin = -10, xmax = 20,
          xtick = {-10,-5,0,5,10,15,20},
          scaled y ticks=false,
          grid = both,
          minor grid style={gray!25},
          major grid style={gray!50},
          legend columns=1,
          legend cell align=left,
          legend style={font=\tiny, at={(0,1)}, anchor=north west},
          legend image post style={line width=0.9pt},
        ]
        \addplot[
        green,% dashed,
        mark=*,
        mark options={solid},
        mark repeat = 1,
        mark size = 2,
        % only marks,
        line width = 1pt,
        forget plot
        ]
        table[x index=0, y index=1, col sep=space]{Figures3/SR_power_1b_1e_2w_4pa_perfect.dat};
        \addplot[
        green,% dashed,
        mark=triangle*,
        mark options={solid},
        mark repeat = 1,
        mark size = 2,
        % only marks,
        line width = 1pt,
        forget plot
        %style = dashed,
        ]
        table[x index=0, y index=1, col sep=space]{Figures3/SR_power_1b_1e_2w_4pa_perfect_no_an.dat};
        \addplot[
        red,
        mark = o,
        mark repeat = 1,
        mark size = 2,
        % only marks,
        line width = 1pt,
        style = solid,
        forget plot
        ]
        table[x index=0, y index=1, col sep=space]{Figures3/SR_power_1b_1e_2w_4pa_imperfect.dat};
        \addplot[
        black,
        mark = o,
        mark repeat = 1,
        mark size = 2,
        line width = 1pt,
        style = solid,
        forget plot
        ]
        table[x index=0, y index=2, col sep=space]{Figures3/SR_power_1b_1e_2w_4pa_imperfect.dat};
        \addplot[
        blue,
        mark = o,
        mark repeat = 1,
        mark size = 2,
        % only marks,
        line width = 1pt,
        style = solid,
        forget plot
        ]
        table[x index=0, y index=3, col sep=space]{Figures3/SR_power_1b_1e_2w_4pa_imperfect.dat};
        \addplot[
        red,
        mark = triangle,
        mark repeat = 1,
        mark size = 2,
        % only marks,
        line width = 1pt,
        style = solid,
        forget plot
        ]
        table[x index=0, y index=1, col sep=space]{Figures3/SR_power_1b_1e_2w_4pa_imperfect_no_an.dat};
        \addplot[
        black,
        mark = triangle,
        mark repeat = 1,
        mark size = 2,
        % only marks,
        line width = 1pt,
        style = solid,
        forget plot
        ]
        table[x index=0, y index=2, col sep=space]{Figures3/SR_power_1b_1e_2w_4pa_imperfect_no_an.dat};
        \addplot[
        blue,
        mark = triangle,
        mark repeat = 1,
        mark size = 2,
        % only marks,
        line width = 1pt,
        style = solid,
        forget plot
        ]
        table[x index=0, y index=3, col sep=space]{Figures3/SR_power_1b_1e_2w_4pa_imperfect_no_an.dat};
        %\addplot[
        %green,
        %mark = o,
        %mark repeat = 1,
        %mark size = 2,
        % only marks,
        %line width = 1pt,
        %style = solid,
        %]
        %table[x index=0, y index=1, col sep=space]{Figures2/SR_power_1b_1e_2w_4pa_panos.dat};
        %\addplot[
        %orange,
        %mark = o,
        %mark repeat = 1,
        %mark size = 2,
        % only marks,
        %line width = 1pt,
        %style = dashed,
        %]
        %table[x index=0, y index=1, col sep=space]{Figures2/SR_power_1b_1e_1w_1pa_OLD.dat};
        \addlegendimage{green, no marks}
        \addlegendentry{Perfect CSI}
        
        \addlegendimage{red, no marks}
        \addlegendentry{$\sigma^2=0.05$}
        
        \addlegendimage{black, no marks}
        \addlegendentry{$\sigma^2=0.15$}
        
        \addlegendimage{blue, no marks}
        \addlegendentry{$\sigma^2=0.3$}
        
        %\addlegendimage{only marks, mark=o, mark options={solid}}
        %\addlegendentry{with AN}
        
        %\addlegendimage{only marks, mark=triangle, mark options={solid}}
        %\addlegendentry{without AN}

        \addlegendimage{empty legend} % draws nothing (defined earlier)
        \addlegendentry{%
          \hspace*{-3.8em}% pull into the (empty) icon slot; adjust to taste
          \tikz[baseline=-0.6ex]{
            \draw[only marks, mark=o, mark options={draw=black, fill=none}]
              plot coordinates {(0,0)};
          }~with AN\quad
          \tikz[baseline=-0.6ex]{
            \draw[only marks, mark=triangle, mark options={draw=black, fill=none}]
              plot coordinates {(0,0)};
          }~without AN
        }
        
    \end{axis}
\end{tikzpicture}
    \caption{SR vs. Transmit Power.}
    \label{fig:SR_power_one}
\end{figure}
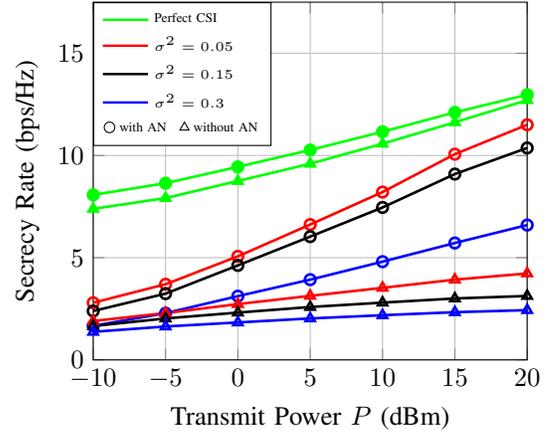

Fig.~\ref{fig:SR_power_one} plots the SR versus the transmit power budget \(P\) for the proposed design, with AN, and the baseline, without AN, under both perfect and imperfect CSI. As expected, SR increases monotonically with \(P\) across all CSI regimes. Although the SINRs of Bob and Eve rise with \(P\), the proposed design achieves a steeper SR slope because the PAs simultaneously reduce Bob’s path loss and align phases, while injecting AN that depresses Eve’s SINR. It is important to note that at sufficiently high \(P\), the imperfect-CSI curves with AN, e.g., \(\sigma^2\in\{0.05,\,0.15\}\), approach the performance level of perfect CSI, underscoring the compensatory effect of AN against channel estimation errors. Hence, while AN offers only modest gains under perfect CSI, it is crucial in realistic imperfect-CSI scenarios, delivering stable SR even under substantial uncertainty in Eve’s channel.

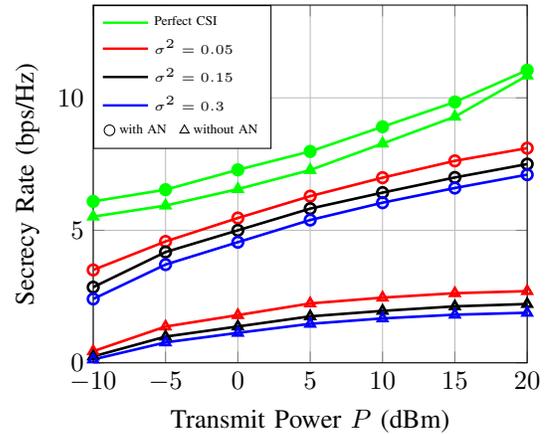
\begin{figure} 
    \centering
    \begin{tikzpicture}
    \begin{axis}[
         width=0.83\linewidth,
          xlabel = {Transmit Power $P$ (dBm)},
          ylabel = {Secrecy Rate (bps/Hz)},
          ymin = 0, ymax = 13.5, xmin = -10, xmax = 20,
          xtick = {-10,-5,0,5,10,15,20},
          scaled y ticks=false,
          grid = both,
          minor grid style={gray!25},
          major grid style={gray!50},
          legend columns=1,
          legend cell align=left,
          legend style={font=\tiny, at={(0,1)}, anchor=north west},
          legend image post style={line width=0.9pt},
        ]
        \addplot[
        green,% dashed,
        mark=*,
        mark options={solid},
        mark repeat = 1,
        mark size = 2,
        % only marks,
        line width = 1pt,
        forget plot
        ]
        table[x index=0, y index=1, col sep=space]{Figures3/SR_power_2b_2e_4w_3pa_perfect.dat};
        \addplot[
        green,% dashed,
        mark=triangle*,
        mark options={solid},
        mark repeat = 1,
        mark size = 2,
        % only marks,
        line width = 1pt,
        forget plot
        %style = dashed,
        ]
        table[x index=0, y index=1, col sep=space]{Figures3/SR_power_2b_2e_4w_3pa_perfect_no_an.dat};
        \addplot[
        red,
        mark = o,
        mark repeat = 1,
        mark size = 2,
        % only marks,
        line width = 1pt,
        style = solid,
        forget plot
        ]
        table[x index=0, y index=1, col sep=space]{Figures3/SR_power_2b_2e_4w_3pa_imperfect.dat}; 
        \addplot[
        black,
        mark = o,
        mark repeat = 1,
        mark size = 2,
        line width = 1pt,
        style = solid,
        forget plot
        ]
        table[x index=0, y index=2, col sep=space]{Figures3/SR_power_2b_2e_4w_3pa_imperfect.dat};
        \addplot[
        blue,
        mark = o,
        mark repeat = 1,
        mark size = 2,
        line width = 1pt,
        style = solid,
        forget plot
        ]
        table[x index=0, y index=3, col sep=space]{Figures3/SR_power_2b_2e_4w_3pa_imperfect.dat};

        \addplot[
        red,
        mark = triangle,
        mark repeat = 1,
        mark size = 2,
        % only marks,
        line width = 1pt,
        style = solid,
        forget plot
        ]
        table[x index=0, y index=1, col sep=space]{Figures3/SR_power_2b_2e_4w_3pa_imperfect_no_an.dat};
        \addplot[
        black,
        mark = triangle,
        mark repeat = 1,
        mark size = 2,
        % only marks,
        line width = 1pt,
        style = solid,
        forget plot
        ]
        table[x index=0, y index=2, col sep=space]{Figures3/SR_power_2b_2e_4w_3pa_imperfect_no_an.dat};
        \addplot[
        blue,
        mark = triangle,
        mark repeat = 1,
        mark size = 2,
        % only marks,
        line width = 1pt,
        style = solid,
        forget plot
        ]
        table[x index=0, y index=3, col sep=space]{Figures3/SR_power_2b_2e_4w_3pa_imperfect_no_an.dat};
        
        %\addplot[
        %green,
        %mark = o,
        %mark repeat = 1,
        %mark size = 2,
        % only marks,
        %line width = 1pt,
        %style = solid,
        %]
        %table[x index=0, y index=1, col sep=space]{Figures2/SR_power_2b_2e_4w_3pa_panos.dat};
        \addlegendimage{green, no marks}
        \addlegendentry{Perfect CSI}
        
        \addlegendimage{red, no marks}
        \addlegendentry{$\sigma^2=0.05$}
        
        \addlegendimage{black, no marks}
        \addlegendentry{$\sigma^2=0.15$}
        
        \addlegendimage{blue, no marks}
        \addlegendentry{$\sigma^2=0.3$}
        
        %\addlegendimage{only marks, mark=o, mark options={solid}}
        %\addlegendentry{with AN}
        
        %\addlegendimage{only marks, mark=triangle, mark options={solid}}
        %\addlegendentry{without AN}

        \addlegendimage{empty legend} % draws nothing (defined earlier)
        \addlegendentry{%
          \hspace*{-3.8em}% pull into the (empty) icon slot; adjust to taste
          \tikz[baseline=-0.6ex]{
            \draw[only marks, mark=o, mark options={draw=black, fill=none}]
              plot coordinates {(0,0)};
          }~with AN\quad
          \tikz[baseline=-0.6ex]{
            \draw[only marks, mark=triangle, mark options={draw=black, fill=none}]
              plot coordinates {(0,0)};
          }~without AN
        }
        
    \end{axis}
    \end{tikzpicture}
    \caption{SR vs. transmit power for 2 Bobs, 2 Eves, 4 waveguides with 3 PAs configuration.}
    \label{fig:SR_power_two}
\end{figure}

In Fig.~\ref{fig:SR_power_two}, the SR is plotted for various values of $P$, however the scenario is for $N=4$, $M=3$, $I=K=2$, to study the multi-Bob and multi-Eve effects, while a comparison is again made with the benchmark without AN, for both perfect and imperfect CSI. Once more, the trend is similar, since the perfect CSI configurations experience higher SRs even for lower power levels, and for imperfect CSI as the power budget increases the \(\sigma^2\in\{0.05,\,0.15, \,0,3\}\) curves of the proposed scheme have an upward trend , while the benchmark achieves notably lower data rates. 
%Another difference here is that the \(\sigma^2=0.3\) line is less effective, showcasing that high channel estimation errors lead to performance degradation when the number of Eves are increased.

\begin{comment}
Fig.~\ref{fig:SR_power_two} shows the multi-user case. SR increases with transmit power in all CSI regimes; the perfect CSI curve is the upper envelope, and imperfect CSI curves are lower as the location-uncertainty variance grows. The green curve is a Bob-centric learning benchmark, where the network is trained to maximize the minimum Bob rate subject to power and spacing constraints, without AN and without any Eve term in the objective. Secrecy is then evaluated a posteriori from the learned beamformers and PA positions. Relative to the perfect CSI curve, the green baseline is consistently lower. Our implementation jointly optimizes the information beams, AN, and PA positions, so it raises Bob’s rate while actively constraining Eve’s, whereas the Bob-only baseline tends to lift both. This confirms that secrecy-aware joint design is essential in PAS in order to have secrecy gains. 
\end{comment}

%Across Fig. ~\ref{fig:SR_power_one} and Fig. ~\ref{fig:SR_power_two}, SR increases monotonically with power, perfect CSI forms the upper envelope, and imperfect CSI curves are pushed downward as the location-uncertainty variance grows. In absolute levels, the single-user setting of Fig.\ref{fig:SR_power_one} attains higher SR than the multi-user setting of Fig.\ref{fig:SR_power_two}. The reasons are structural, because with two Bobs the transmitter must split power and manage inter-user interference, which reduces each Bob’s SINR. With two Eves and a worst-case objective, the design faces a tighter eavesdropping bottleneck, effectively elevating their SINR. 

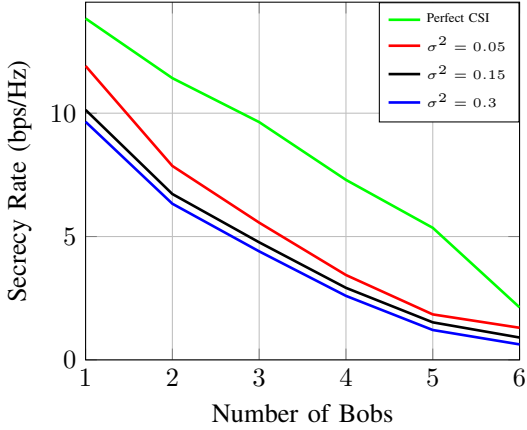
\begin{figure} 
    \centering
    \begin{tikzpicture}
    \begin{axis}[
        width=0.83\linewidth,
        xlabel = {Number of Bobs},
        ylabel = {Secrecy Rate (bps/Hz)},
        ymin = 0,
        ymax = 14.5,
        xmin = 1,
        xmax = 6,
        % ytick = {1.0,1.1,1.2,1.3,1.4,1.5},
        %xtick = {-10,-5,0,5,10,15,20},         
        % },
        scaled y ticks=false, % Prevents multiplication by powers of 10
        grid = both,
        minor grid style={gray!25},
        major grid style={gray!50},
        legend columns=1, 
    legend entries ={Perfect CSI, $\sigma^2=0.05$, $\sigma^2=0.15$, $\sigma^2=0.3$},
        legend cell align = {left},
        legend style={font=\tiny},
        legend style={at={(1,1)},anchor=north east},
        legend image post style={scale=0.7}, % Adjust the line size in the legend
        % legend pos = north west
        ]
        \addplot[
        green,
        mark repeat = 1,
        mark size = 2,
        % only marks,
        line width = 1pt,
        style = solid,
        ]
        table[x index=0, y index=1, col sep=space]{Figures3/SR_Bobs_1e_6w_4pa_perfect.dat};
        \addplot[
        red,
        mark repeat = 1,
        mark size = 2,
        line width = 1pt,
        style = solid,
        ]
        table[x index=0, y index=1, col sep=space]{Figures3/SR_Bobs_1b_1e_6w_4pa_imperfect.dat};
        \addplot[
        black,
        mark repeat = 1,
        mark size = 2,
        % only marks,
        line width = 1pt,
        style = solid,
        ]
        table[x index=0, y index=2, col sep=space]{Figures3/SR_Bobs_1b_1e_6w_4pa_imperfect.dat};
        \addplot[
        blue,
        mark repeat = 1,
        mark size = 2,
        % only marks,
        line width = 1pt,
        style = solid,
        ]
        table[x index=0, y index=3, col sep=space]{Figures3/SR_Bobs_1b_1e_6w_4pa_imperfect.dat};
    \end{axis}
\end{tikzpicture}
    \caption{SR vs. number of Bobs with $P=10$ dBm.}
    \label{fig:Bobs}
\end{figure}

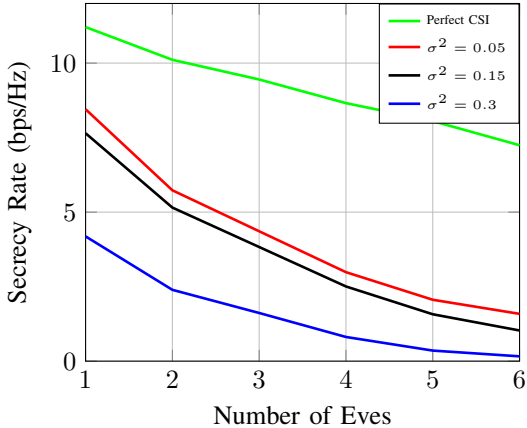
\begin{figure} 
    \centering
    \begin{tikzpicture}
    \begin{axis}[
        width=0.83\linewidth,
        xlabel = {Number of Eves},
        ylabel = {Secrecy Rate (bps/Hz)},
        ymin = 0,
        ymax = 12,
        xmin = 1,
        xmax = 6,
        % ytick = {1.0,1.1,1.2,1.3,1.4,1.5},
        %xtick = {-10,-5,0,5,10,15,20},         
        % },
        scaled y ticks=false, % Prevents multiplication by powers of 10
        grid = both,
        minor grid style={gray!25},
        major grid style={gray!50},
        legend columns=1, 
    legend entries ={Perfect CSI, $\sigma^2=0.05$, $\sigma^2=0.15$, $\sigma^2=0.3$},
        legend cell align = {left},
        legend style={font=\tiny},
        legend style={at={(1,1)},anchor=north east},
        legend image post style={scale=0.7}, % Adjust the line size in the legend
        % legend pos = north west
        ]
        \addplot[
        green,
        mark repeat = 1,
        mark size = 2,
        % only marks,
        line width = 1pt,
        style = solid,
        ]
        table[x index=0, y index=1, col sep=space]{Figures3/SR_Eves_1b_2w_4pa_perfect.dat};
        \addplot[
        red,
        mark repeat = 1,
        mark size = 2,
        line width = 1pt,
        style = solid,
        ]
        table[x index=0, y index=1, col sep=space]{Figures3/SR_Eves_1b_1e_2w_4pa_imperfect.dat};
        \addplot[
        black,
        mark repeat = 1,
        mark size = 2,
        % only marks,
        line width = 1pt,
        style = solid,
        ]
        table[x index=0, y index=2, col sep=space]{Figures3/SR_Eves_1b_1e_2w_4pa_imperfect.dat};
        \addplot[
        blue,
        mark repeat = 1,
        mark size = 2,
        % only marks,
        line width = 1pt,
        style = solid,
        ]
        table[x index=0, y index=3, col sep=space]{Figures3/SR_Eves_1b_1e_2w_4pa_imperfect.dat};
    \end{axis}
\end{tikzpicture}
    \caption{SR vs. number of Eves with $P=10$ dBm.}
    \label{fig:Eves}
\end{figure}

Figs. \ref{fig:Bobs} and \ref{fig:Eves} demonstrate how the number of Bobs and Eves affects the achieved SR. In both figures, SR decreases monotonically, as expected, since on the one hand when the number of Bobs increases, the available resources are split to serve all of them, while on the other hand, more Eves increase the overall eavesdropping capabilities, creating a challenging environment for maintaining stable SR performance. SR degradation is steeper when the number of Bobs increases because the power must be split across multiple information streams, and inter-user interference grows. In contrast, the presence of more Eves leads to a more gradual decline since resources focus on serving Bob while protecting him from eavesdropping. The transmitter can focus power toward Bob and reallocate AN and beam directions to contain the additional Eves' SINR without sharing power with more Bobs. As uncertainty variance grows, the protection must cover a wider region, becoming less targeted, thus the SR reduction becomes more noticeable. However, it is important to note that, with perfect CSI, the performance of the proposed scheme remains almost stable and slowly degrades with increasing numbers of Eves. 

\begin{figure} 
    \centering
    \begin{tikzpicture}
      \begin{axis}[
        view={0}{90},                % top‑down
        width=6.5cm, height=6.5cm,
        xlabel={$x$ [m]},
        ylabel={$y$ [m]},
        title={Secrecy‑rate map},
        colorbar,
        colormap/viridis,
        point meta min=0,
        point meta max=11.28525,
        enlargelimits=false,
        axis on top,
      ]
    \addplot3[
      surf,
      shader=interp,    % smooth interpolation
      patch type=bilinear,
      draw=none,
      mesh/rows=51
    ]
    table[x=x,y=y,z=z,col sep=space,header=true]{Figures3/data_new_2.csv};
      \end{axis}
    \end{tikzpicture}
    \caption{SR map as a function of Bob’s position under perfect CSI with $P=10$ dBm.}
    \label{fig:3d}
\end{figure}
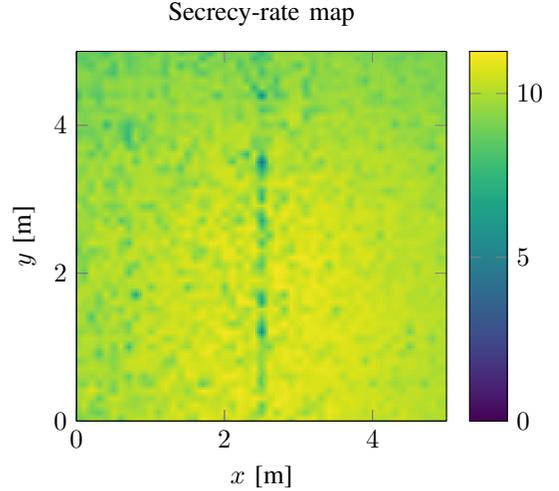

Fig.~\ref{fig:3d} shows a 2D heatmap of the achieved SR when Bob moves over a dense grid in the room with a configuration of two waveguides, each with four PAs. For each location of Bob, we solve the perfect CSI problem and record the resulting SR in the presence of two Eves, placed at $(2.5, 1.25)$ and $(2.5, 3.75)$. First, lower SR values are observed in regions near the Eves. This behavior is expected, since more spatial and power resources must be spent for AN to limit the leakage when Bob approaches Eve. Lower SR is also observed along the line connecting the two Eves. This is due to the geometry of the problem, as steering high gain to Bob while simultaneously suppressing leakage toward both Eves along that direction reduces the available spatial DoF, producing degraded SRs in these positions. Additionally, SR tends to decrease slightly near the room boundaries compared to the interior. This is because, when Bob is at the edge of the room, the PA ensemble can mostly reach him from one side. With fewer distinct directions available, it is harder to give Bob a strong advantage over the Eves nearby, so SR values are lower near the edges than in more central positions.

\section{Conclusion}\label{sec:Conc}

In this paper, we introduced an AN-aided beamforming scheme for secure downlink transmissions in PASs. An optimization problem was formulated that jointly optimizes the beamforming vector, the AN covariance matrix, and the positions of the PAs to maximize the SR. We studied the effects of both perfect and imperfect CSI for the eavesdroppers by modeling location errors as an ellipsoidal uncertainty set to provide a more realistic representation of the latter. We also derived a closed-form solution for the single waveguide scenario with one user. To address the challenges of multi-user and multi-waveguide scenarios, each of which is equipped with multiple PAs, we proposed a DNN-aided joint optimizer that can achieve near-optimal solutions while satisfying the problem's constraints. Finally, numerical results demonstrate that our scheme improves the SR of PAS baselines across single- and multi-user settings under both perfect and imperfect CSI, while the gains persisted for various layouts and parameter values.

\appendices

\section{Proof of Lemma \ref{l1}}\label{ap:ferrari}
Since the objective function in \eqref{eq:opt_single_x} depends only on the scalar variable $\tilde{x}_{P}$, we differentiate it with respect to $\tilde{x}_{P}$ and set the derivative equal to zero to identify possible extrema. After some algebraic manipulation, the resulting first-order derivative is expressed as
\begin{equation}\label{eq:derivative_first}
\begin{aligned}
\frac{d\,\mathrm{SR}}{d\tilde{x}_{P}} =& \frac{\tilde{x}_{P}-x_b}{(\tilde{x}_{P}^2-2x_b \tilde{x}_{P}+K_2)^2+K_1(\tilde{x}_{P}^2-2x_b\tilde{x}_{P}+K_2)} \\
-&\frac{\tilde{x}_{P}-x_e}{(\tilde{x}_{P}^2-2x_e \tilde{x}_{P}+K_3)^2+K_1(\tilde{x}_{P}^2-2x_e\tilde{x}_{P}+K_3)},
\end{aligned}
\end{equation}
where $K_1=\lVert w\rVert^2 \eta/\sigma_B^2$, $K_2=x_b^2+y_b^2+d^2+\eta R_m/\sigma_B^2$ and $K_3=x_e^2+y_e^2+d^2+\eta R_m/\sigma_E^2$.
By setting \eqref{eq:derivative_first} equal to zero and performing some algebraic manipulations, we obtain the following quartic equation:
\begin{equation}\label{eq:derivative_pin}
\begin{aligned}
&\frac{d\,\mathrm{SR}}{d\tilde{x}_{P}} = 0 \;\Rightarrow \alpha_4\,\tilde{x}_{P}^4+\alpha_3\,\tilde{x}_{P}^3+\alpha_2\,\tilde{x}_{P}^2+\alpha_1\,\tilde{x}_{P}-\alpha_0=0, 
\end{aligned}
\end{equation}
with the coefficients of $\tilde{x}_{P}$ defined as 
\begin{align}
\alpha_4\,&=\,3 x_e - 3 x_b\nonumber,\\
\alpha_3\,&=\,4\,x_b^2 \;-\; 4x_e^2 \;+\; 2K_2 \;-\; 2K_3\nonumber,\\
\alpha_2\;&=\;K_1 x_e \;-\; K_1 x_b \;-\; 4K_2 x_b \;+\; 2K_3 x_b \nonumber\\
\;&-\; 2K_2 x_e \, +\; 4K_3 x_e \;+\; 4x_e^2 x_b \;-\; 4x_b^2 x_e,\\
\alpha_1\;&=\;K_2^2 \;-\; K_3^2 \;+\; K_1K_2 \;-\; K_1K_3 \nonumber\\
\;&+\; 4K_2x_bx_e \;-\; 4K_3x_bx_e\nonumber,\\
\alpha_0\;&=\; K_2^2 x_e \;+\; K_3^2 x_b \;+\; K_1K_3 x_b \;-\; K_1K_2 x_e.\nonumber
\end{align}
To simplify \eqref{eq:derivative_pin}, we perform the substitution  $\tilde{x}_{P}=\tilde{u}_{P}-\alpha_3/(4\alpha_4)$, which effectively eliminates the cubic term. Consequently, the equation can be rewritten in its depressed form as
\begin{equation}\label{eq:depressed}
    \tilde{u}_{P}^4+\alpha_2'\,\tilde{u}_{P}^2+\alpha_1'\,\tilde{u}_{P}+\alpha_0'=0,
\end{equation}
where the coefficients of $\tilde{u}_{P}$ are given as
\begin{align}
&\alpha_2'\,=\,\frac{\alpha_2}{\alpha_4}-\frac{3}{8}{\left(\frac{\alpha_3}{\alpha_4}\right)}^2,\\
&\alpha_1'\,=\,\frac{\alpha_1}{\alpha_4}-\frac{\alpha_2\alpha_3}{2\alpha_4^2}+\frac{1}{8}{\left(\frac{\alpha_3}{\alpha_4}\right)}^3,\\
&\alpha_0'\;=\;\frac{\alpha_0}{\alpha_4}+\frac{\alpha_3^2\alpha_2}{16\alpha_4^3}-\frac{\alpha_3\alpha_1}{4\alpha_4^2}-\frac{3}{256}{\left(\frac{\alpha_3}{\alpha_4}\right)}^4.
\end{align}
To solve \eqref{eq:depressed}, we need to factorize the quartic polynomial as the product of two quadratic terms, i.e.,
\begin{equation}\label{eq:factor}
    \tilde{u}_{P}^4+\alpha_2'\tilde{u}_{P}^2+\alpha_1'\tilde{u}_{P}+\alpha_0'=(\tilde{u}_{P}^2+p_1\tilde{u}_{P}+p_0)\,(\tilde{u}_{P}^2+q_1\tilde{u}_{P}+q_0).
\end{equation}
From the factorization in~\eqref{eq:factor}, matching the coefficients of each power of $\tilde{u}_P$ on both sides yields the following system of equations:
\begin{subequations}
\begin{align}
-p_1 = q_1&,\\
p_0q_0 = \alpha_0'&,\label{eq:pq}\\
p_1(q_0-p_0) = \alpha_1'&,\\
p_0 + q_0 - p_1^2 = \alpha_2'&.
\end{align}
\end{subequations}
Since solving this system directly is challenging, we introduce the substitution $\omega = q_0 - p_0.$ With this substitution, the system can be reformulated as 
\begin{subequations}
\begin{align}
p_1 =& -q_1 = \frac{\alpha_1'}{\omega}, \\
p_0 =& \frac{1}{2}\left(\alpha_2' - \omega + \frac{\alpha_1'^2}{\omega^2}\right), \label{eq:p0}\\
q_0 =& \omega + \frac{1}{2}\left(\alpha_2' - \omega + \frac{\alpha_1'^2}{\omega^2}\right). \label{eq:q0}
\end{align}
\end{subequations}
Substituting \eqref{eq:p0} and \eqref{eq:q0} into the \eqref{eq:pq} gives an equation for $m$, i.e.,
\begin{equation}
    \omega^6+\beta_2\omega^4+\beta_1 \omega^2+\beta_0=0,
\end{equation}
where $\beta_2=4\,\alpha_0'-\alpha_2'^2$, $\beta_1=-2\alpha_2' \alpha_1'^2$ and $\beta_0=-\alpha_1'^4$. Setting $l = \omega^2$, the equation simplifies to the following cubic form:
\begin{equation}
    l^3+\beta_2l^2+\beta_1l+\beta_0=0.
\end{equation}
To eliminate the quadratic term, we introduce the substitution $l=z-\beta_2/3$, which transforms the polynomial into
\begin{equation}\label{eq:cardano}
    z^3+\beta_1'z+\beta_0'=0,
\end{equation}
where $\beta_1'=\beta_1+\beta_2^2/3$ and $\beta_0'=2/(3\beta_2)^3-\beta_1\beta_2/3+\beta_0$. It is noted that \eqref{eq:cardano} is now in the depressed cubic form and can be solved using Cardano's formula. In this formulation, the discriminant is defined as $\Delta=\left(\beta_0'/2\right)^2+\left(\beta_1'/3\right)^3$, which indicates the nature of the roots. The real solutions for $z$ are given by
\begin{equation}
z=\left\{
\begin{aligned}
&\sqrt[3]{-\frac{\beta_0'}{2}+\sqrt{\Delta}} + \sqrt[3]{-\frac{\beta_0'}{2}-\sqrt{\Delta}},\quad \Delta\geq0,\\
&2\sqrt{-\frac{\beta_1'}{3}}\cos{\left(\frac{\theta+2\pi k}{3}\right)}, \, k=0,1,2,\quad \Delta<0,
\end{aligned}
\right.
\end{equation}
where 
\begin{equation}
    \theta=\arccos{\left(\frac{-\frac{\beta_0'}{2}}{\sqrt{-\left(\frac{\beta_1'}{3}\right)^3}}\right)}.
\end{equation}
It should be noted that although the case $\Delta < 0$ yields three distinct real roots, a single real solution for $z$ is sufficient for our purposes to facilitate the factorization of $\tilde{u}_{P}$. Once $z$ is determined, we reverse the substitutions to recover $l$ and then $\omega$, with only one value of $\omega$ required to complete the factorization in \eqref{eq:factor}. Next, we solve the two resulting quadratic equations from the second part of \eqref{eq:factor} to obtain their roots and substitute these values to determine the candidate values of $\tilde{x}_{P}$. Finally, we select the value that maximizes our objective function while satisfying the waveguide boundary constraint $\mathrm{C}_1$, which completes the proof.

\section{Proof of Lemma \ref{l2}}\label{ap:monotonic}

The function in \eqref{eq:SR_fixed_x} depends on a single variable, $R_m$. To evaluate its monotonicity, we differentiate the function for $R_m$. After algebraic manipulations, we obtain the derivative as
\begin{equation}
    \frac{d\,\mathrm{SR}}{dR_m} =  \frac{\eta\,\left(r_B^2\sigma_B^2 - r_E^2\sigma_E^2\right)}{(\eta R_m + r_B^2\sigma_B^2)(\eta R_m + r_E^2\sigma_E^2)}.
\end{equation}
Since $\eta$ and the denominator are strictly positive, the sign of the derivative depends only on the term $r_B^2\sigma_B^2 - r_E^2\sigma_E^2$, which is independent of $R_m$. Consequently, the derivative keeps a constant sign, and the monotonicity of $\mathrm{SR}$ is characterized by
\begin{equation}
\left\{
\begin{aligned}
r_B^2\sigma_B^2 > r_E^2\sigma_E^2 &\quad \Longrightarrow \quad \mathrm{SR}\uparrow, \\
r_B^2\sigma_B^2 < r_E^2\sigma_E^2 &\quad \Longrightarrow \quad \mathrm{SR}\downarrow.
\end{aligned}
\right.
\end{equation}
Since the derivative has a constant sign, the function has no stationary points. Thus, any extremum must occur at one of the endpoints of $R_m$, namely at $0$ and $P$, which completes the proof.

\bibliographystyle{IEEEtran}
\bibliography{references}

\end{document}